\documentclass[acmtog,nonacm]{acmart}
\acmSubmissionID{215}

\usepackage{booktabs} 

\usepackage[ruled]{algorithm2e} 

\SetAlFnt{\small}
\SetAlCapFnt{\small}
\SetAlCapNameFnt{\small}
\SetAlCapHSkip{0pt}

\acmJournal{TOG}

\usepackage{amsmath}
\usepackage{amsfonts}
\usepackage{multirow}
\usepackage{xurl}
\usepackage{adjustbox}
\usepackage{tabularx}
\usepackage{overpic}
\usepackage{subcaption}
\newcommand{\ignore}[1]{}

\newcommand{\FLIP}{\protect{\reflectbox{F}LIP}\xspace}

\begin{document}

\title{Scalar Spatiotemporal Blue Noise Masks}

\author{Alan Wolfe}
\affiliation{%
  \institution{NVIDIA}
  \country{USA}
  }
\email{awolfe@nvidia.com}
\author{Nathan Morrical}
\affiliation{%
  \institution{NVIDIA and University of Utah}
  \country{USA}
}
\email{nmorrical@nvidia.com}

\author{Tomas Akenine-M\"oller}
\affiliation{%
  \institution{NVIDIA}
  \country{Sweden}
}
\email{takenine@nvidia.com}

\author{Ravi Ramamoorthi}
\affiliation{%
  \institution{NVIDIA and UC San Diego}
  \country{USA}
}
\email{ravir@cs.ucsd.edu}


\begin{abstract}
  Blue noise error patterns are well suited to human perception, and when applied 
  to stochastic rendering techniques, blue noise masks (blue noise textures) minimize
  unwanted low-frequency noise in the final image. Current methods of applying blue noise masks at
  each frame independently produce white noise frequency spectra temporally. This white 
  noise results in slower integration convergence over time and unstable results 
  when filtered temporally. Unfortunately, achieving temporally stable blue 
  noise distributions is non-trivial since 3D blue noise does not exhibit the 
  desired 2D blue noise properties, and alternative approaches degrade the 
  spatial blue noise qualities.
  We propose novel blue noise patterns that, when animated, produce values at a 
  pixel that are well distributed over time, converge rapidly for Monte Carlo 
  integration, and are more stable under TAA, while still retaining spatial blue noise
  properties. To do so, we propose an extension to the well-known \emph{void and
  cluster} algorithm that reformulates the underlying energy function to produce
  spatiotemporal blue noise masks. 
  These masks exhibit blue noise frequency spectra in {\em both the spatial and 
  temporal domains}, resulting in visually pleasing error patterns, rapid 
  convergence speeds, and increased stability when filtered temporally. We 
  demonstrate these improvements on a variety of applications, including 
  dithering, stochastic transparency, ambient occlusion, and volumetric 
  rendering.
  By extending spatial blue noise to spatiotemporal blue noise, we overcome the 
  convergence limitations of prior blue noise works, enabling new applications 
  for blue noise distributions.
\end{abstract}  

%
%



\begin{teaserfigure}
  \includegraphics[width=\linewidth]{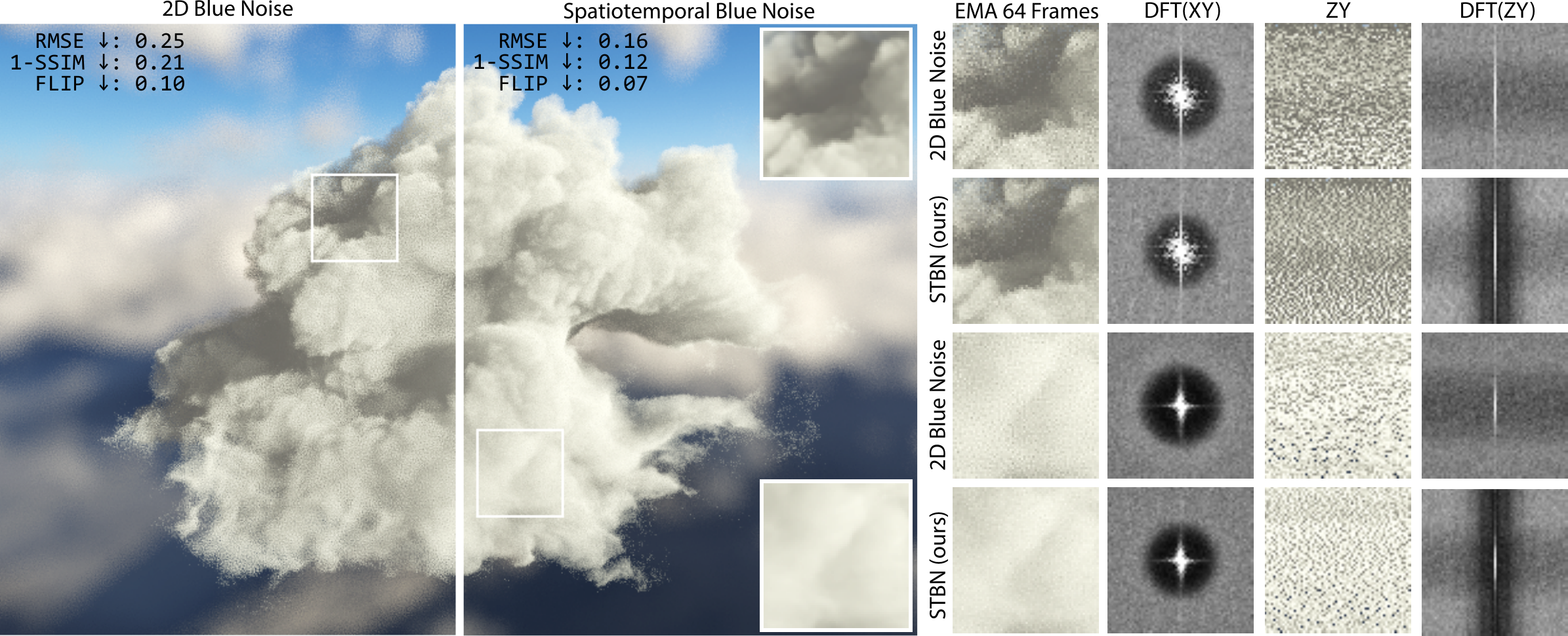}
 \caption{All images rendered using an exponential moving average (EMA) with $\alpha=0.1$. On the left is an image of the
 Disney Cloud~\shortcite{Disney2020} rendered using stochastic single scattering, where free-flight distances are sampled 
 using a series of blue noise masks over time. Traditional 2D blue noise masks (far left) are easy to filter spatially, 
 but exhibit a white noise signal over time, making the underlying signal difficult to filter temporally. Our spatiotemporal 
 blue noise (STBN) masks (right of large image) additionally exhibit blue noise in the temporal dimension, resulting in a 
 signal that is easier to filter over time. On the far right, we show two crops of the main image, as well as their 
 corresponding discrete Fourier transforms over both space (DFT(XY)) and time (DFT(ZY)). 
 The Z axis is time. The ground truth is shown in the insets in the large image (upper and lower right corners). 
}
 \label{fig_teaser}
\end{teaserfigure}

\maketitle


\section{Introduction}
\label{sec:intro}

\begin{figure}[tb]
	\centering
	\setlength{\fboxsep}{0pt}%
	\setlength{\fboxrule}{0.0mm}%
	\setlength{\tabcolsep}{1.5pt}%
	\renewcommand{\arraystretch}{0.95}
	\newcommand{\imgW}{0.24\columnwidth}
	{\scriptsize 
		\begin{tabular}{cccc}
			\includegraphics[width=\imgW]{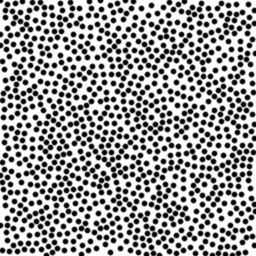} &
			\includegraphics[width=\imgW]{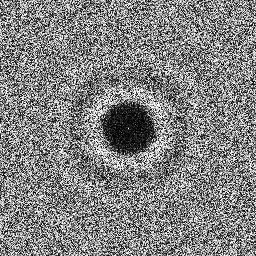} &
			\includegraphics[width=\imgW]{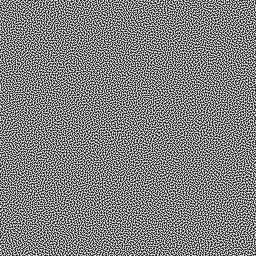} &
			\includegraphics[width=\imgW]{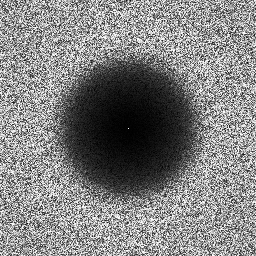} \\
			Blue Noise Samples & DFT Magnitude & Blue Noise Mask & DFT Magnitude
		\end{tabular}
	}
	\caption{The left two images show blue noise sample points in 2D and the magnitude of the discrete Fourier transform (DFT).
	These are in contrast to the two images to the right, which show a blue noise mask and the magnitude of the DFT. In this case, the mask is a 2D image
	where each pixel stores a single ``random value'' with blue noise properties.
	Both samples and masks show attenuated low frequencies but have different uses. Our work focuses on masks.}
	\vspace*{-.2in}
	\label{fig_sample_patterns_vs_masks}
\end{figure}


Blue noise error distributions have long been recognized as an appealing alternative to purely random white noise distributions in computer graphics. 
In real-time and complex stochastic rendering scenarios, sample counts per frame are constrained. 
As a result, many modern visual effects depend on amortizing sampling expense over space and time to achieve higher quality images at an acceptable performance. 
If the rendered frames produce a white noise error distribution, the resulting images will contain difficult to filter low-frequency clusters. 
Alternatively, blue noise patterns contain only high-frequency noise that results in a perceptually uniform, cluster-free distribution. 
However, few attempts have been made to extend spatial blue noise patterns to produce easier to filter distributions along the temporal domain.

In addition to filtering spatially, current real-time techniques often use temporal antialiasing (TAA)~\cite{Yang2020} to filter these distributions over time. 
If samples over time were also made to follow a blue noise distribution, TAA could produce similar quality results with a lower alpha blend value, reducing temporal lag, or more accurate results at the same alpha value.
Similarly, techniques may seek to integrate over multiple samples per pixel while still maintaining blue noise error properties spatially. 
Computer displays---and even human perception---can perform some amount of implicit integration over time, especially at high frame rates~\cite{Andersson2019}. 
These situations motivate the need for good sampling patterns over time in addition to space. 

There has been considerable effort to generate good blue noise patterns---made popular in rendering by Mitchell~\shortcite{Mitchell91}, but also in error diffusion, ordered dithering, and digital halftoning by both Ulichney~\shortcite{Ulichney1993} and Mitsa and Parker~\shortcite{Mitsa91}. 
These methods can broadly be divided into two categories: those that generate a discrete set of blue noise distributed \textit{sample points}~\cite{Balzer08,Hongwei10,deGoes2012}, and those that generate continuous values in an image~\cite{Ulichney1993,Georgiev2016},
commonly referred to as blue noise \textit{masks}. 
See Fig.~\ref{fig_sample_patterns_vs_masks} for comparison. 
Blue noise masks differ from blue noise sample sets in
that masks provide a scalar value for a given 2D pixel location, while sample sets return a 2D vector a given 1D index.
In this paper, we focus on the former, generating spatiotemporal blue noise masks. 

\begin{figure}[tb]
	\centering
	\setlength{\fboxsep}{0pt}%
	\setlength{\fboxrule}{0.0mm}%
	\setlength{\tabcolsep}{1.5pt}%
	\renewcommand{\arraystretch}{1.}
	\newcommand{\imgW}{0.49\columnwidth}
	{\scriptsize 
		\begin{tabular}{cc}
			White Noise & Blue Noise \\
			\includegraphics[width=\imgW,trim=0 1cm 0 1cm,clip]{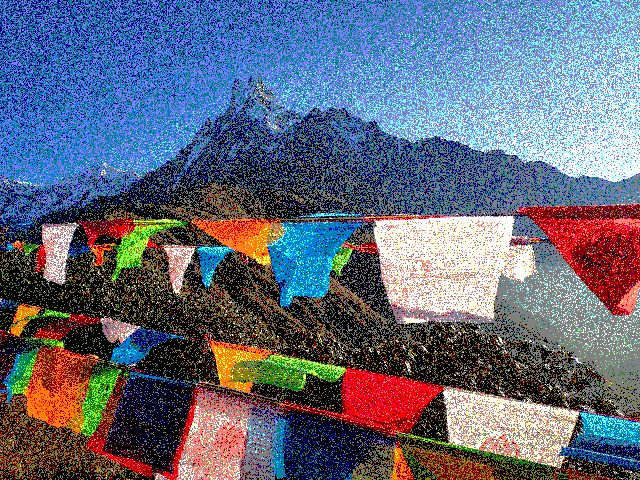} &
			\includegraphics[width=\imgW,trim=0 1cm 0 1cm,clip]{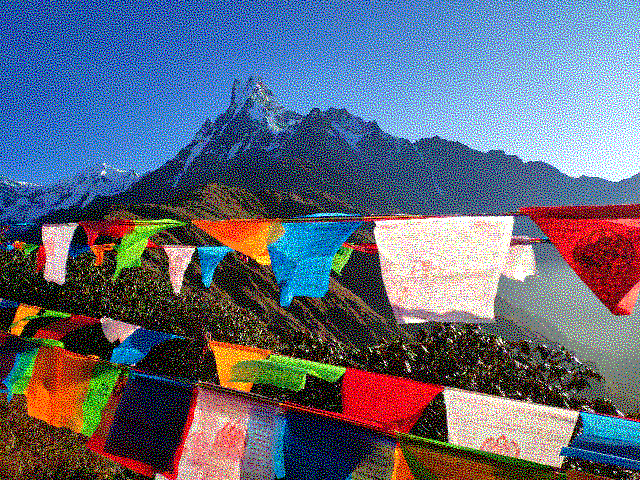} \\
			White Noise + Gaussian Blur & Blue Noise + Gaussian Blur\\
			\includegraphics[width=\imgW,trim=0 1cm 0 1cm,clip]{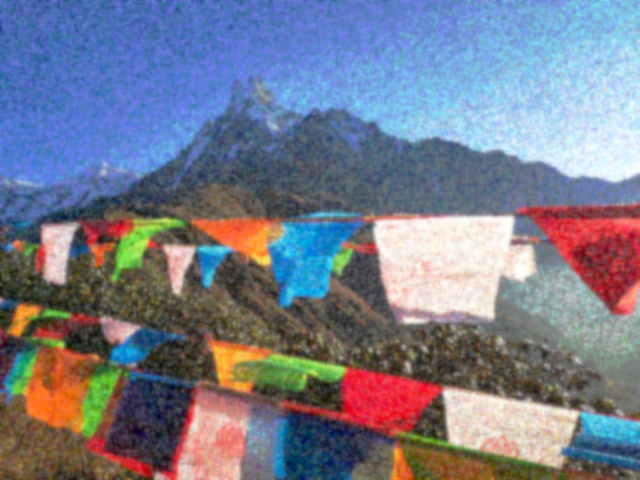} &
			\includegraphics[width=\imgW,trim=0 1cm 0 1cm,clip]{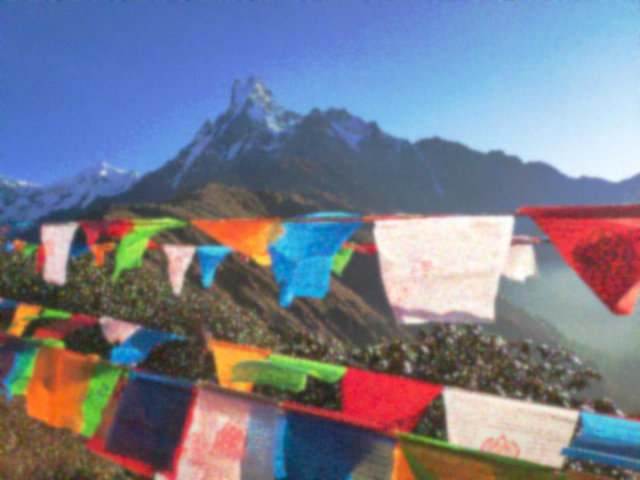} \\
		\end{tabular}
	}
    \vspace*{-.1in}
	\caption{A comparison of white noise versus blue noise, and corresponding spatially filtered results. White noise
	results in visual clumps when filtered with a low-pass filter (blur) due to low frequencies present in the noise.
	Blue noise images do not contain these low frequencies, and thus do not contain clumps in the filtered output. }
    \vspace*{-.1in}
	\label{fig_noise_filter}
\end{figure}

Blue noise masks can be used by stochastic rendering algorithms as pseudorandom number generators in order to produce 
perceptually uniform noise in the resulting image.
These high-frequency screen-space error patterns can then be relatively easily removed using one or more low-pass filters.
Additionally, since a local neighborhood of pixels following a blue noise distribution contains a more diverse sampling of the underlying function,  
spatially filtered blue noise images more closely approximate the ground truth.
Figures~\ref{fig_teaser} and~\ref{fig_noise_filter} show these effects visually. Our paper extends blue noise masks
to account for the temporal domain as well. 

Previous works such as INSIDE~\cite{Gjoel2016} and by Heitz and Belcour~\shortcite{Heitz2019} address the time axis by using a low discrepancy sequence to spatially offset a blue noise texture every frame.
This causes the frequency spectrum to be approximately white noise over time due to subsequent texture samples being far apart and thus uncorrelated. 
Other works use an array of two-dimensional blue noise masks, sampling from different masks over time. 
However, this too produces white noise distributions along the temporal axis, since these blue noise textures were generated independently. 
Using a three-dimensional blue noise mask texture provides neither two-dimensional spatial blue noise in each time slice, nor blue noise properties along the time dimension~\cite{Peters2017}.
Another approach is to add the golden ratio to each pixel value in a 2D blue noise texture every frame to make each pixel value use the golden ratio additive recurrence low discrepancy sequence on the time axis.
This improves sampling over time at the cost of damaging the frequency content of the spatial domain.
The image quality and frequency spectra of these cases can be seen in Fig.~\ref{fig_3dmask_dfts}.

What is really desired is a technique where each two-dimensional mask has blue noise properties, and where each pixel has good one-dimensional sampling properties over the time dimension.
These masks could then be used by denoising algorithms like TAA~\cite{Yang2020} and SVGF~\cite{Schied2017,Schied2018} to produce higher quality, more temporally stable results. 
To the best of our knowledge, we present the first algorithm for spatiotemporal blue noise mask generation.
Our contributions include:
\begin{itemize}
	\item An algorithm that generates spatiotemporal 
		blue noise masks by
		modifying the well known void and cluster algorithm~\cite{Ulichney1993} in Section~\ref{sec:algorithm}.
	\item Practical analysis of our algorithm's frequency spectrum and convergence speeds in
		Sections~\ref{sec_stbn_analysis} and~\ref{sec_qualities_taa}.
	\item Using spatiotemporal masks to generate spatiotemporal point sets in Section~\ref{sec_pointsets}.
	\item Evaluations of our spatiotemporal masks when applied to a variety of techniques in Section~\ref{sec_applications}.
 	\item Theoretical frequency analysis in Appendix~\ref{sec_freq_analysis}.
	\item Extension to higher dimensions in Appendix~\ref{higher_dimensional_masks}.
\end{itemize}

\section{Previous Work}
\label{sec_prevwork}

The most prevalent method for generating blue noise masks is the \emph{void and cluster} algorithm by Ulichney~\shortcite{Ulichney1993}. 
This algorithm works by repeatedly inserting values into the largest voids between pixels in a sparse image until all pixels are filled in. 
Using this approach, the void and cluster algorithm is able to generate high-quality blue noise masks, works in any dimension, and allows tuning of a sigma parameter to control the frequencies present in the mask.  
Void and cluster also generates blue noise such that, when a threshold is applied, sparse pixels are  arranged in a blue noise sampling pattern. 
This property can be useful for stippling applications, such as importance map-driven sparse sampling or stochastic alpha.

Georgiev and Fajardo~\shortcite{Georgiev2016} showed how these blue noise masks can be applied to Monte Carlo ray tracing, and demonstrated impressive results.  
These authors also introduced an algorithm to generate blue noise masks by starting with a white noise mask and swapping pixels to reduce an energy function.  
Where void and cluster only stores a scalar value per pixel entry, the blue noise masks by Georgiev and Fajardo store an entire vector value per pixel entry.
However, the later vector valued blue noise masks lose the thresholding property of void and cluster if using the algorithm to generate single scalar masks.
The algorithm also solves the problem less directly than void and cluster---doing random swaps to try to make the texture more like a blue noise mask, and
using simulated anealing to try to avoid local minima.

Despite the good spatial properties of these masks, if multiple such masks are used over time, values along the temporal dimension produce undesirable white noise.
For the void and cluster masks, this white noise is caused by each generated mask being independent of the next.
For the vector valued masks by Georgiev and Fajardo, this temporal white noise is caused by vectors only being swapped spatially, but not values within the per-pixel vectors.
Additionally, Georgiev and Fajardo note that the overall Monte Carlo error when using these masks is the same as if using white noise, despite looking noticeably less noisy.  

Heitz and Belcour~\shortcite{Heitz2019} point out that the improvements of blue noise dithered sampling are quickly lost as sample counts and sample dimensionality increase.
They propose an alternative a posteriori method that reorganizes nearby per-pixel random seeds such that the corresponding rendered image approximately follows a blue noise pattern.
This process is done by sorting a small region of pixels from both a rendered image and a blue noise mask by respective luminance, then matching these sorted pixel sets together to permute the
rendering seeds for the next frame. They also allow improvement frame over frame by using a precomputed permutation to move the seeds into the desired configutation of the next frame.
The benefit of this a posteriori reformulation is that the rendering function is treated as a black box, where only the initial random seeds are modified.
As a result, any state of the art rendering method can be used during rendering without modification while still achieving a spatial blue noise distribution.
A downside of this algorithm is that pixel seeds are always sorted into a target noise pattern, which can deviate from the desired ground truth image pattern. 
This a posteriori algorithm is meant primarily for use with spatial denoising, but the authors do also consider the temporal dimension.
Heitz and Belcour animate a single target blue noise mask over time by using the generalized golden ratio R2 sequence~\cite{Roberts2018} to offset the screen space
position of the texture.  This makes the noise be blue over space, but uncorrelated white noise over time, as shown in Fig.~\ref{fig_bn_autocorrelation}.
Observing that this technique could instead target our spatiotemporal blue noise textures, we analyze it in more detail in Appendix~\ref{sec_freq_analysis} and
show some results targeting our blue noise.

More relevant to our work is that by Ahmed and Wonka~\shortcite{Ahmed2020}, where the authors propose a method that
uses a well-converging sampling sequence while simultaneously resulting in a screen-space blue noise error pattern.  
This is done by using a locality-preserving mapping of 2D pixel coordinates into a 1D pixel sequence, and then using 1D low discrepancy sequences across those pixels.  
In this work, the axis of time is considered in the form of progressive sampling, but ultimately this approach results in a trade-off between quality over space or time.

The work by Heitz et al.~\shortcite{Heitz2019B} also aims to use a faster converging sequence.
Their proposed method uses a precomputed permutation table that attempts to preserving a spatial blue noise pattern while simultaneously using the Sobol sequence.  
This permutation table is made by starting with white noise seeds and swapping pixels to minimize an energy function.  
The technique provides multiple samples per pixel, which could be distributed over the time dimension, but ultimately this comes at the cost of spatial image quality as shown in Fig.~\ref{fig_3dmask_dfts}.
Also shown in Fig.~\ref{fig_3dmask_dfts}, this Sobol sequence strobes over time. 
For offline rendering, this strobing only occurs during the convergence process, and therefore does not effect the final converged frame; 
however, for real time rendering this strobing results in undesirable temporal instability.

A study of blue noise for real-time rendering is presented by Gjoel and Svendsen~\shortcite{Gjoel2016}, which deals with a plethora of subjects including triangular distributed noise, animating blue noise, and using blue noise in various techniques.  
Similar to Heitz and Belcour's work~\shortcite{Heitz2019}, a low discrepancy sequence is used to offset where the texture is read every frame to animate the blue noise, which results in white noise over time.

Another method for animating blue noise can be found in the post by Wolfe~\shortcite{Wolfe2017} which adds the golden ratio to a blue noise mask every frame.
This causes each pixel to use the golden ratio additive recurrence over time for improved convergence speeds, but damages blue noise frequencies spatially
as shown in Fig~\ref{fig_3dmask_dfts} and shows a similar strobing pattern over time as Heitz and Belcour~\shortcite{Heitz2019B}. 
Other rank 1 lattices~\cite{Keller2008} could be used instead of the golden ratio additive recurrence, but these lattices also result in damaged blue noise frequencies.
This damage is caused by different locations of the blue noise texture rolling over from high to low values at differing rates, which fundamentally alters the frequency the texture.

Blue noise masks are tailored toward low sample counts, but are not the only way to achieve high quality
results with low sample counts. For instance, high quality, low sample count ray traced direct lighting
is achieved by Bitterli et al.~\shortcite{Bitterli2020} without the use of blue noise.
\section{Background: Void And Cluster Algorithm}
\label{sec:voidandcluster}

We base our algorithm on the void and cluster method~\cite{Ulichney1993}, which we extend to handle
the spatiotemporal domain in Section~\ref{sec:algorithm}. There has not been much advancement in
generating high quality blue noise masks since the void and cluster algorithm, apart from the work
by Georgiev and Fajardo~\cite{Georgiev2016} which adds vector values to the masks, but is inapropriate
for scalar value use.

Here, we review the basic void and cluster algorithm. 
To generate a blue noise mask $M$ of dimension $D$, the algorithm stores a boolean per pixel specifying 
whether the pixel is turned \emph{on} (emits energy to the energy field) and an integer index per pixel 
specifying the order that this pixel was turned on in. This ordering index is used to compute the final 
output color for that pixel, where the first pixel to be turned on is black, and the last pixel to be 
turned on is white.
Every pixel $\mathbf p$,  
which is turned on, gives energy to every point $\mathbf q$
in the energy field using:
\begin{equation}
	\label{eqn_vcenergy}
	E(\mathbf p,\mathbf q) = \exp{\left(-\frac{\Vert \mathbf p-\mathbf q \Vert ^2}{2\sigma^2}\right)},
\end{equation}
where $\mathbf p$ and $\mathbf q$ are the integer coordinates and distances are computed on wrapped boundaries,
i.e., \textit{toroidal} wrapping.
The $\sigma$ is a tuneable parameter, which controls energy falloff over distance, and thus frequency content.
Ulichney~\shortcite{Ulichney1993} recommends
$\sigma=1.5$.
This is a Gaussian blur, which is used to distribute the energy from activated pixels into
the energy field over distance.

The energy field $E$ is discretized onto a grid the same size as the mask $M$ and is defined as:
\begin{equation}
	\label{eqn_vcenergyfull}
	F(\mathbf p) = \sum_{\mathbf q \in M} E(\mathbf p,\mathbf q).
\end{equation}

\paragraph*{Initial Binary Pattern Generation}
The first step in the void and cluster algorithm is to generate an initial \textit{binary} pattern
where less than or equal to half of the pixels are turned on.  This can be done using white noise or
nearly any other pattern.
These pixels need to be transformed into a blue noise distribution before proceeding to the next step. 
That transformation is done by repeatedly turning off the tightest cluster pixel - defined as $\sup_{\mathbf p \in M} F(\mathbf p)$ -
and turning on the largest void pixel - defined as $\inf_{\mathbf p \in M} F(\mathbf p)$.
This process is repeated until the same pixel is found for both operations.  Once this condition is met, the
algorithm will have converged, and the initial binary pattern will be blue noise distributed. Once these pixels
have been assigned an on or an off state, they can be assigned an ordering, which will occur in the next step.

%
%

\paragraph*{Phase I - Initial Pattern Ordering}
Ordering is done by iteratively turning off the current tightest cluster pixel, and assigning that pixel an \emph{ordering}, 
which is found by counting how many pixels remain on after turning off the current tightest cluster pixel. 
This process is repeated until all pixels are turned off. Once all pixels are off, the initial binary pattern is turned back on.
When paired with the ordering generated earlier, this initial binary pattern of pixels describes a sparse binary sequence that 
follows a spatial blue noise distribution.

\paragraph*{Phase II - Order First Half of Pixels}
At the end of \emph{Phase I}, a small subset of pixels are on and ordered. However, the remaining pixels are off and 
unordered. In \emph{Phase II}, these remaining off pixels are turned on, one at a time, until half of the pixels are 
turned on. Similar to initialization, these pixels are turned on in order from largest void pixel to smallest void pixel.
The ordering given to each pixel turned on is set to be the number of pixels that were on before this pixel was turned on.
At the end of this operation, half of the pixels will be on and ordered to follow a blue noise sequence.

\paragraph*{Phase III - Order Second Half of Pixels}

Once half of our pixels are ordered, in \emph{Phase III} the state of all the pixels are reversed. Pixels that are on 
are turned off, and vice versa. Next, we follow a similar process to \emph{Phase II}, and iteratively find the current 
largest valued cluster pixel that is on and turn that pixel off, assigning to that pixel an order found by counting the 
number of pixels that were off before this pixel was turned off. When all pixels are turned off, \emph{Phase III} is 
finished and all pixels will have an ordering.

\paragraph*{Texture Finalization}
After all pixels are ordered, this ordering is used to compute the final pixel values in the output image. For floating-point 
images, per-pixel values can be found by dividing the order of each pixel by the total number of pixels in the image. For
$k$-bit images, these pixel values must be remapped to 0 to $2^k-1$. 


\section{Spatiotemporal Blue Noise Masks}
\label{sec_algorithm}
As mentioned in Section~\ref{sec:intro}, our work aims to extend purely spatial blue noise masks to also exhibit a blue 
noise distribution along an additional third temporal dimension. This can be used 
to amortize sampling 
expense over time and converge these samples progressively, improving final image quality in the process. 

In Section~\ref{sec:algorithm}, we present a novel algorithm for generating spatiotemporal blue noise masks using a modified 
version of the void and cluster algorithm. Then, in Sections~\ref{sec_stbn_analysis} and~\ref{sec_qualities_taa},
we perform a thorough analysis of the 
resulting spatiotemporal blue noise masks.
From our spatiotemporal blue noise masks, spatiotemporal point sets can be generated and are discussed in Section~\ref{sec_pointsets}.
Finally, extensions of spatiotemporal blue noise masks to higher dimensions 
are discussed in Appendix~\ref{higher_dimensional_masks}.

\subsection{Spatiotemporal Void and Cluster Algorithm} \label{sec:algorithm}
Our ultimate goal with spatiotemporal blue noise is to have each pixel within the mask vary over time to enable temporal integration. 
At the same time, we must preserve the spatial blue noise patterns in our mask to still easily filter our signal spatially for any point in time. 
One approach would be to generate several independent 2D blue noise textures, one per frame. These masks would be easy to filter spatially; 
however, pixels would exhibit a white noise distribution over time that would be difficult to filter temporally. 
Instead, we aim to generate multiple two-dimensional blue noise masks where each pixel individually follows a one-dimensional blue noise distribution over time. 


To this end, we reformulate the void and cluster algorithm from Section~\ref{sec:voidandcluster} such that noise generation is driven by a novel energy function, in the spirit of Equation~\ref{eqn_vcenergy}.
Instead of only distributing energy in two dimensions, we distribute energy in three dimensions, and constrain this energy function to achieve our desired spatial and temporal blue noise properties.
When evaluating this three-dimensional energy function, if the given pixel coordinates belong to the same two-dimensional layer, we return the same two-dimensional energy as in Equation~\ref{eqn_vcenergy}.
Otherwise, if the given pixel coordinates match spatially but belong to different two-dimensional layers, we distribute energy temporally. 
Finally, if the given pixel coordinates differ in all dimensions, we do not distribute energy (i.e., we return an energy of zero).
\begin{figure}[tb]
	\centering
	\includegraphics[width=1.0\columnwidth]{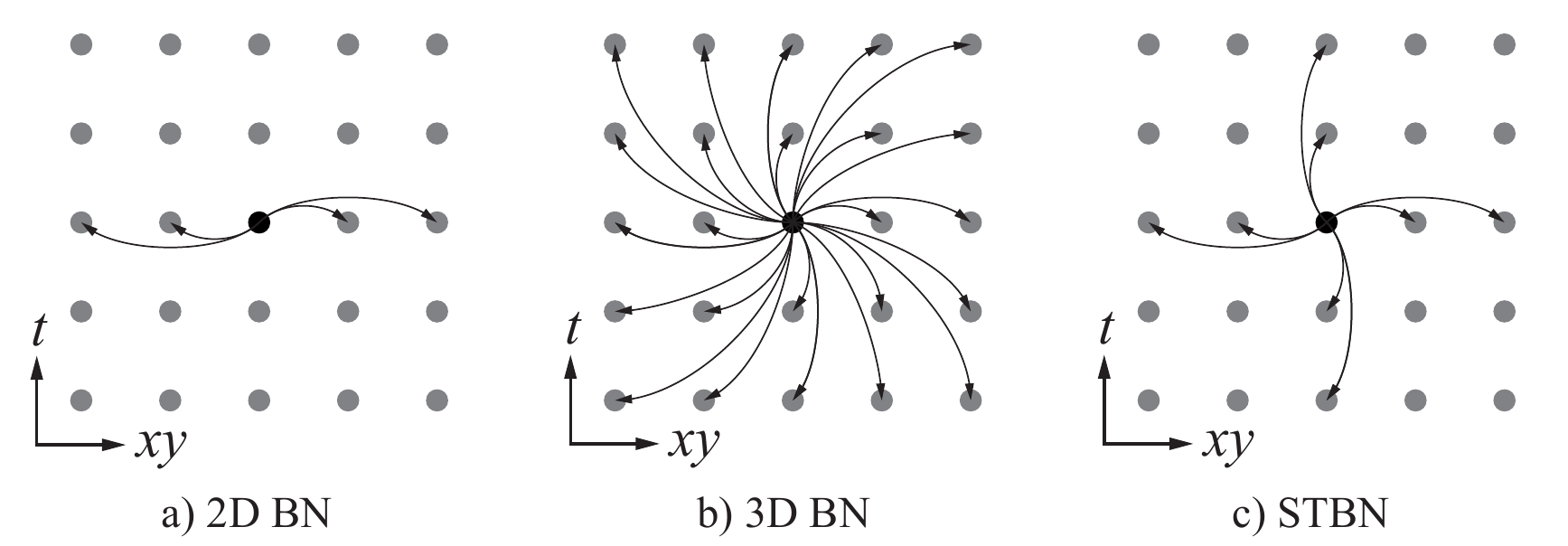} 
	\caption{An illustration of void-and-cluster energy evaluation, where we have compressed the spatial $xy$
	dimensions to a single horizontal dimension and time is on the vertical axis.
	2D blue noise measures energy to all pixels in the same 2D layer,
	while 3D blue noise measures energy to all pixels in the entire 3D texture.
	Our spatiotemporal blue noise on the right measures distance in the 2D layer 
	and along the time dimension for the current pixel.
	}
	\label{fig_energy_function_evaluation_directions}
	\vspace{-1em}
\end{figure}

By ensuring the first spatial condition, we guarantee that the pattern generated by void and cluster will follow a blue noise distribution spatially for each layer of our mask,
and by ensuring the second temporal condition, each pixel in our mask will exhibit blue noise properties over time.
This energy function is illustrated in Figure~\ref{fig_energy_function_evaluation_directions}.


We can formulate this three-dimensional energy function as follows, where
a pixel in the three-dimensional spatiotemporal blue noise texture is denoted
as $\mathbf p =(\mathbf p_{xy}, p_z)=(p_x,p_y,p_z)$.
Our modified energy formulation is then
\begin{equation}
\label{eqn_vcenergy_new}
E(\mathbf p,\mathbf q) =
\begin{cases}
	\exp{\left(-\frac{\Vert \mathbf p_{xy}-\mathbf q_{xy} \Vert ^2}{2\sigma_{xy}^2}\right)}, & \text{if\ } p_z = q_z \\
	\exp{\left(-\frac{( p_z-q_z )^2}{2\sigma_z^2}\right)}, &  \text{if\ } \mathbf p_{xy} = \mathbf q_{xy}\\
0, &\text{otherwise.} \\
\end{cases}
\end{equation}

When generating spatiotemporal blue noise, we had best results using an initial binary pattern
density of 10\% of the pixels, and we use $\sigma = 1.9$ for all axes. 
It is important to note that the distance used by Equation~\ref{eqn_vcenergy_new} is computed
toroidally on all axes, so that individual texture slices tile well over space, and
the temporal qualities also tile well over time. 

\begin{figure}[tb]
	\centering
	\setlength{\fboxsep}{0pt}%
	\setlength{\fboxrule}{0.0mm}%
	\setlength{\tabcolsep}{1.5pt}%
	\renewcommand{\arraystretch}{0.95}
	\newcommand{\imgW}{0.3\columnwidth}
	{\scriptsize 
		\begin{tabular}{l|ccc}
			& Mask & DFT(XY) & DFT(XZ) \\ 
			\hline
			\rotatebox{90}{$\textbf{STBN (ours)}$} &
			\includegraphics[width=\imgW]{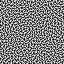} &
			\includegraphics[width=\imgW]{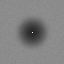} &
			\includegraphics[width=\imgW]{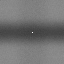} \\
			\hline
			\rotatebox{90}{$\text{2DBN}$} &
			\includegraphics[width=\imgW]{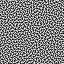} &
			\includegraphics[width=\imgW]{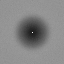} &
			\includegraphics[width=\imgW]{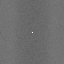} \\
			\hline
			\rotatebox{90}{$\text{3DBN}$} &
			\includegraphics[width=\imgW]{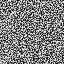} &
			\includegraphics[width=\imgW]{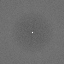} &
			\includegraphics[width=\imgW]{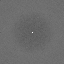} \\
			\hline
			\rotatebox{90}{GR: $\text{2D frame 34}$} &
			\includegraphics[width=\imgW]{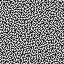} &
			\includegraphics[width=\imgW]{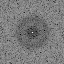} &
			\includegraphics[width=\imgW]{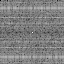} \\
			\hline
			\rotatebox{90}{$\text{HB LDBN Frame 24}$} &
			\includegraphics[width=\imgW]{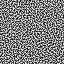} &
			\includegraphics[width=\imgW]{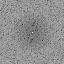} &
			\includegraphics[width=\imgW]{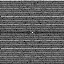} \\
			\hline
		\end{tabular}
	}
	\caption{Analysis of various blue noise masks.  Only our spatiotemporal blue noise (STBN) has blue
	noise over space and better than white noise over the $z$-axis (time).  The DFTs are averaged to show
	expected frequency spectra, except for golden ratio (GR) animated blue noise and Heitz/Belcour (HB LDBN),
	which highlights ways they damage spatial frequencies at specific frame numbers.
 	}
	\vspace*{-.2in}
	\label{fig_3dmask_dfts}
\end{figure}
\begin{figure*}
	\centering
	\setlength{\fboxsep}{0pt}%
	\setlength{\fboxrule}{0.0mm}%
	\setlength{\tabcolsep}{1.5pt}%
	\renewcommand{\arraystretch}{0.95}
	\newcommand{\imgW}{0.7\columnwidth}
	{\scriptsize 
		\begin{tabular}{ccc}
		\includegraphics[width=\imgW]{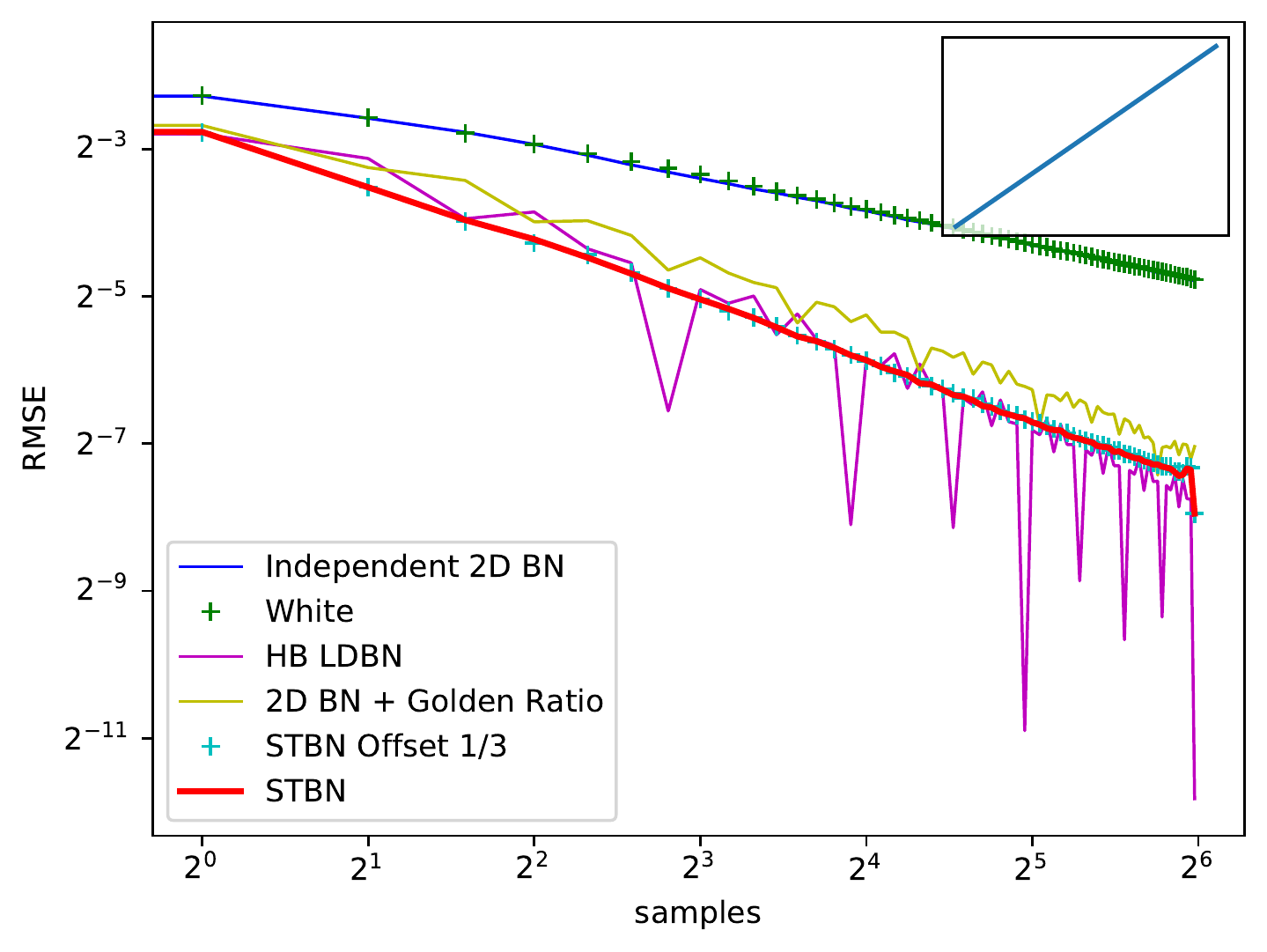} &
		\includegraphics[width=\imgW]{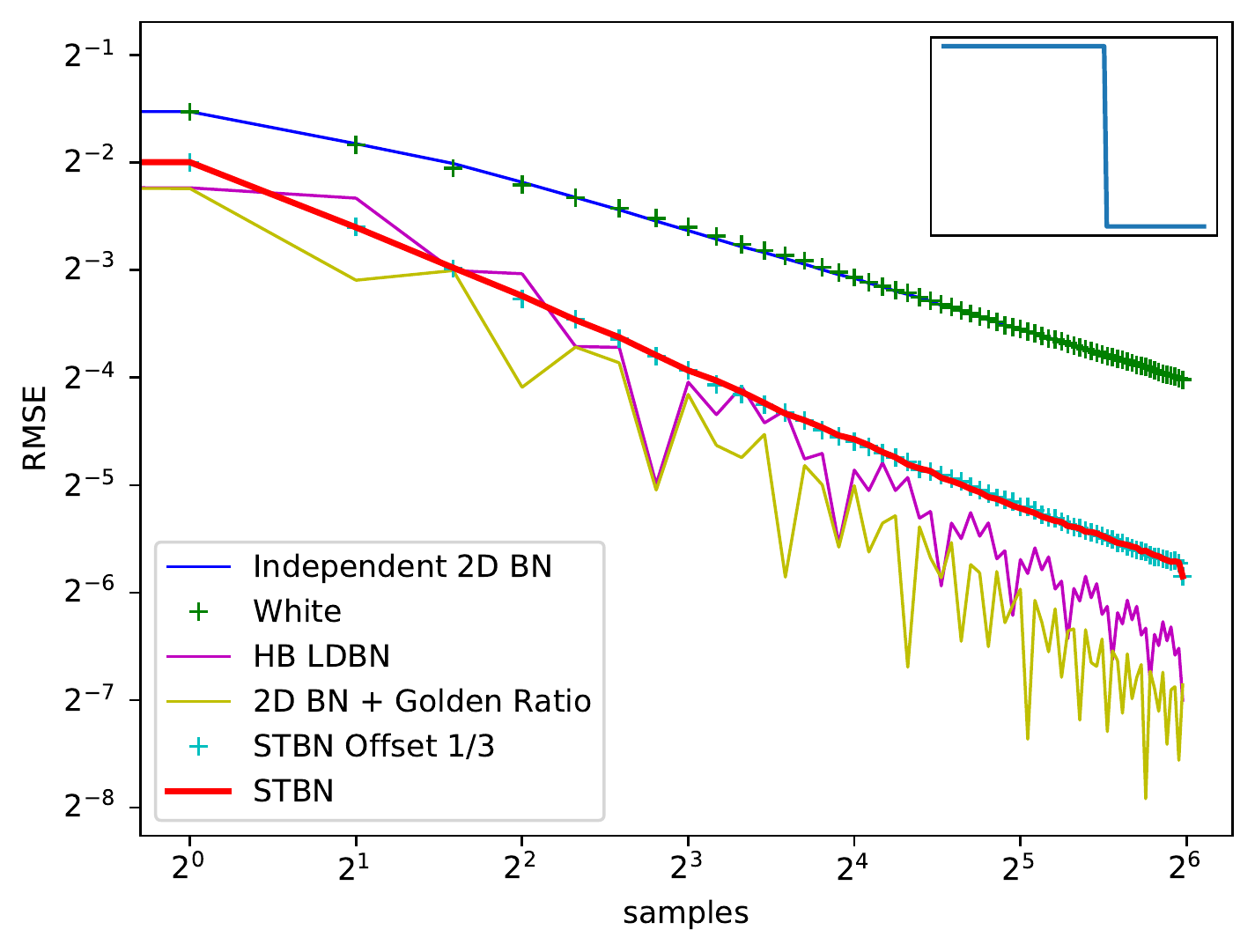} &
		\includegraphics[width=\imgW]{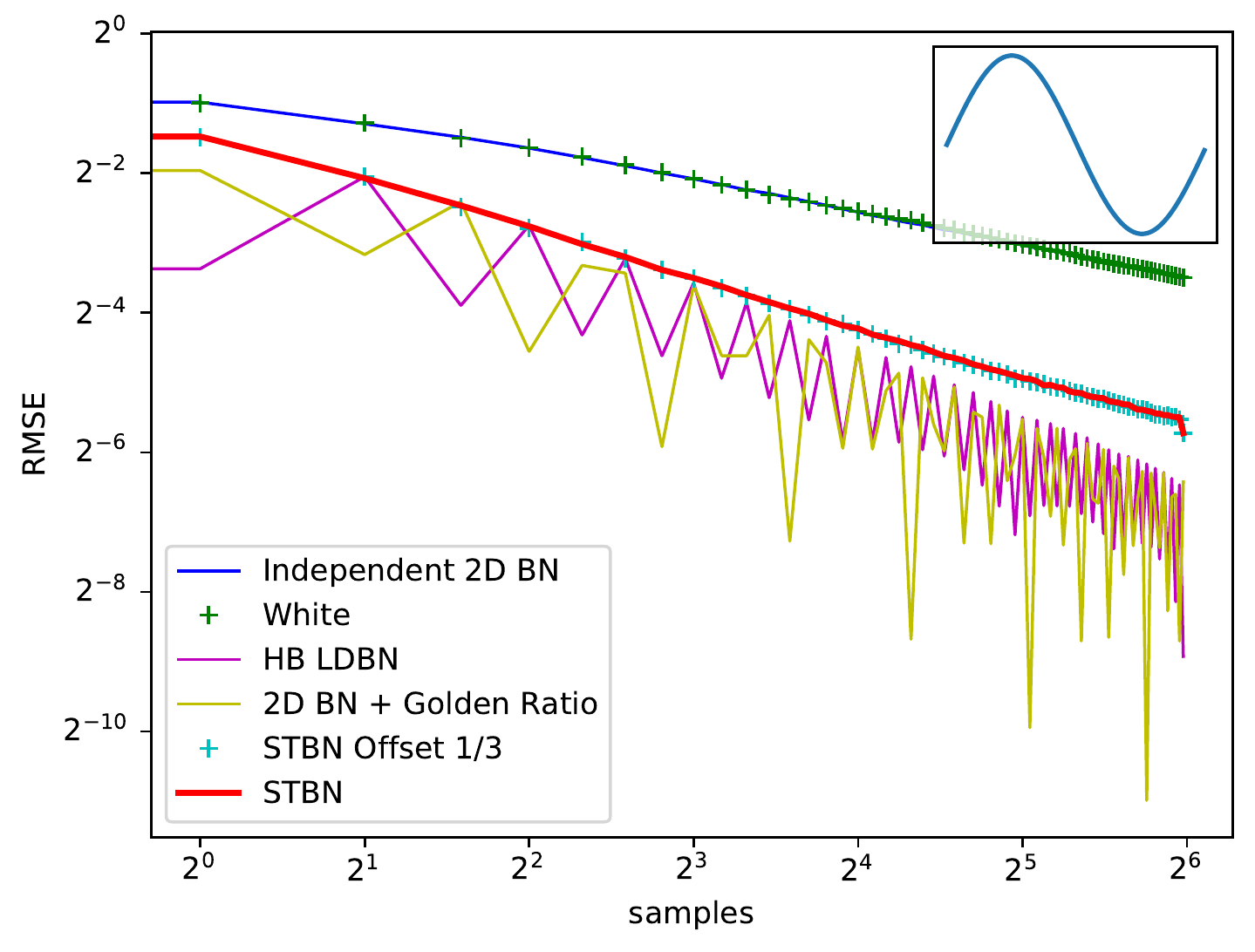} \\
		\end{tabular}
	}
	\vspace*{-.2in}
    \caption{RMSE for sampling ramp, step, and sine functions with 64 samples, combined with Monte
	Carlo integration. Independent slices of 2D blue noise show that they are white noise over time.
	STBN shows how it is toroidally progressive when offsetting it by 1/3 of its index count shows
	no difference in convergence. Blue noise animated by the golden ratio and Heitz and Belcour (HB LDBN)
	low discrepancy blue noise often show good convergence but are erratic which causes per-pixel 
    strobing when viewed in motion.}
    \vspace*{-.1in}
	\label{fig_3dmask_convergence_mc}
\end{figure*}

\subsection{Spatiotemporal Blue Noise Analysis}
\label{sec_stbn_analysis}
To analyze the results from our algorithm, we generated a spatiotemporal blue noise (STBN) mask of $64^3$ resolution using the algorithm in Section~\ref{sec:algorithm}.
In addition, we generated a $64^3$ 3D blue noise mask (3DBN) and 64 independent 2D blue noise masks (2DBN) of size $64^2$ using the void and cluster algorithm in Section~\ref{sec:voidandcluster}.  
We also made a spatiotemporal mask by using a single 2D blue noise mask, then adding the golden ratio to the values of this mask each frame for 63 frames (GR), wrapping these values around 0 to 1. 
Finally, we generated a $64^3$ blue noise mask using the low discrepancy sampler by Heitz and Belcour~\shortcite{Heitz2019}.
Comparative results demonstrating noise patterns and corresponding discrete Fourier transforms (DFT) are shown in Fig.~\ref{fig_3dmask_dfts}.

Our goal in this work is to obtain a noise pattern that exhibits blue noise properties in each spatial 2D slice, as well as a blue noise pattern along the time axis.
These blue noise patterns can then be filtered spatially and temporally for use in low sample count real-time applications.
As can be seen in Figure~\ref{fig_3dmask_dfts}, only our method provides those two features simultaneously.

As predicted, 2D blue noise exhibits the desired spatial blue noise properties, but generates white noise over time (as shown on the vertical axis of DFT(XZ)). 
Somewhat counterintuitively, 3D blue noise exhibits low quality blue noise properties both spatially and temporally. 
On further analysis, this result makes sense. 3D blue noise generates a dark sphere in a 3D DFT rather than a dark circle in a 2D DFT.
As a result, $xy$ cross sections over $z$ contain dark circles of differing radii, and therefore generate differing blue noise qualities over time, from white noise to blue noise and back to white noise.
A deeper analysis of 3D blue noise can be found from Peters~\shortcite{Peters2017}.

\begin{figure}[tb]
	\centering
	\setlength{\fboxsep}{0pt}%
	\setlength{\fboxrule}{0.0mm}%
	\setlength{\tabcolsep}{0.0pt}%
	\renewcommand{\arraystretch}{0.95}
	\newcommand{\imgW}{0.16\columnwidth}
	{\scriptsize 
		\begin{tabular}{cccccc}
			GR DFT(XZ) & GR 0 & GR 13 & GR 14 & GR 21 & GR 22 \\
			\includegraphics[width=\imgW]{Images/BlueNoise/2DGR/DFTSlices/XZ/34.mag.png} &
			\includegraphics[width=\imgW]{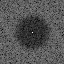} &
			\includegraphics[width=\imgW]{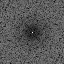} &
			\includegraphics[width=\imgW]{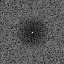} &
			\includegraphics[width=\imgW]{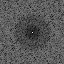} &
			\includegraphics[width=\imgW]{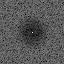} \\
            STBN DFT(XZ) & STBN 0 & STBN 13 & STBN 14 & STBN 21 & STBN 22 \\
            \includegraphics[width=\imgW]{Images/BlueNoise/ST/DFTSlices/XZ.mag.png} &
            \includegraphics[width=\imgW]{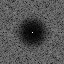} &
			\includegraphics[width=\imgW]{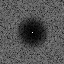} &
			\includegraphics[width=\imgW]{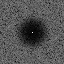} &
			\includegraphics[width=\imgW]{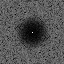} &
			\includegraphics[width=\imgW]{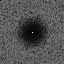} \\
		\end{tabular}
	}
	\caption{A comparison of blue noise quality between golden radio (GR) blue noise and STBN (ours) over time. The blue noise qualities of GR 
	are periodically damaged over time, resulting in a strobing pattern. STBN results in consistent blue noise 
	qualities from one frame to the next.}
	\vspace*{-.2in}
	\label{fig_lds_strobing}
\end{figure}

Both the golden ratio and the low discrepancy sequence by Heitz and Belcour~\shortcite{Heitz2019} exhibit damaged blue noise frequencies spatially, at the trade-off of better convergence over time.
However, both of these low discrepancy noise patterns exhibit strobing patterns along the time dimension,
which can be seen as distinct horizontal lines in DFT(XZ). See also Figure~\ref{fig_lds_strobing}.
For offline rendering, these strobing patterns only occur during convergence, and therefore do not affect the final frame.
For real-time applications though, this temporal strobing will lead to a more challenging signal to filter temporally.

\begin{figure*}
	\centering
	\setlength{\fboxsep}{0pt}%
	\setlength{\fboxrule}{0.0mm}%
	\setlength{\tabcolsep}{1.5pt}%
	\renewcommand{\arraystretch}{0.95}
	\newcommand{\imgW}{0.7\columnwidth}
	{\scriptsize 
		\begin{tabular}{ccc}
		\includegraphics[width=\imgW]{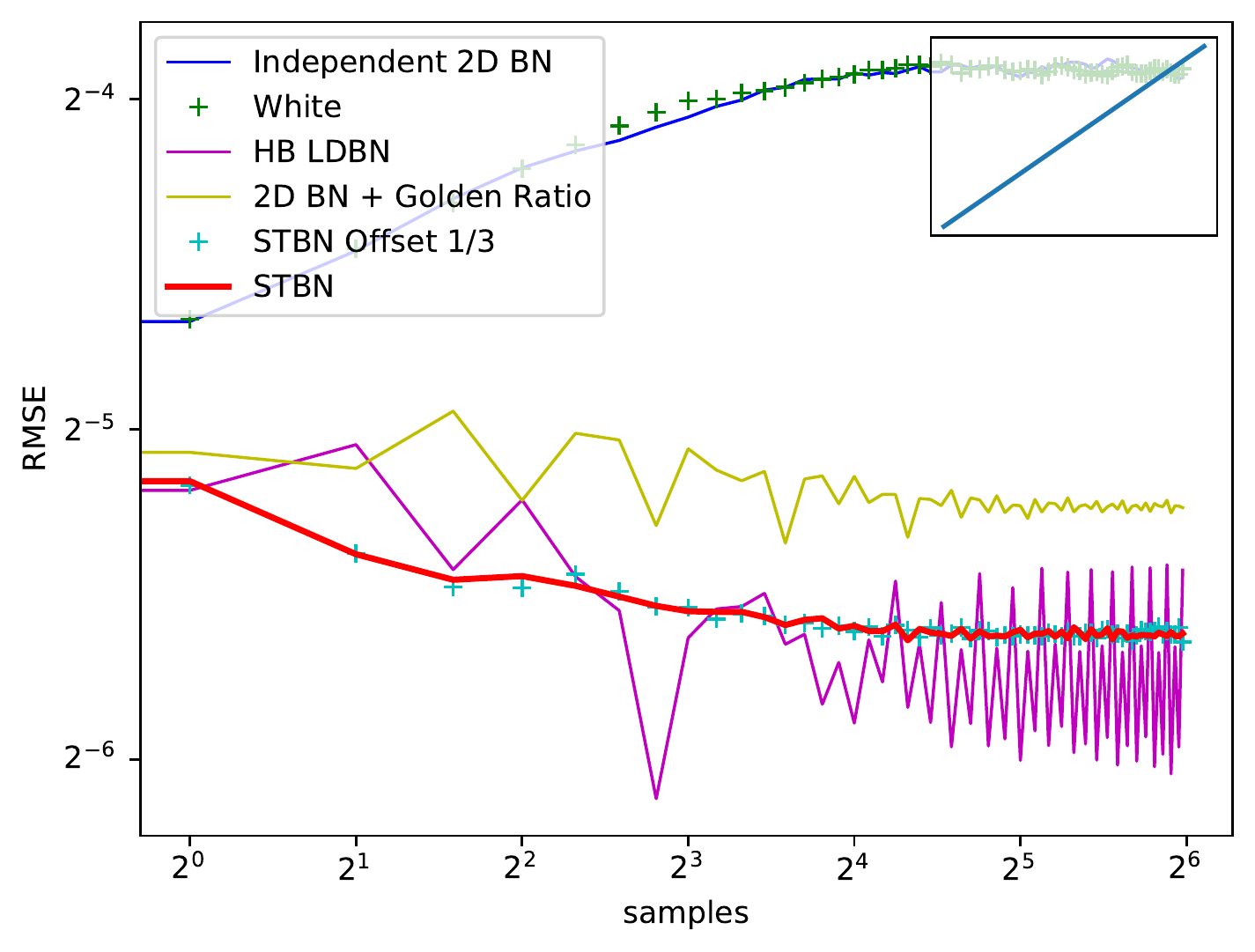} &
		\includegraphics[width=\imgW]{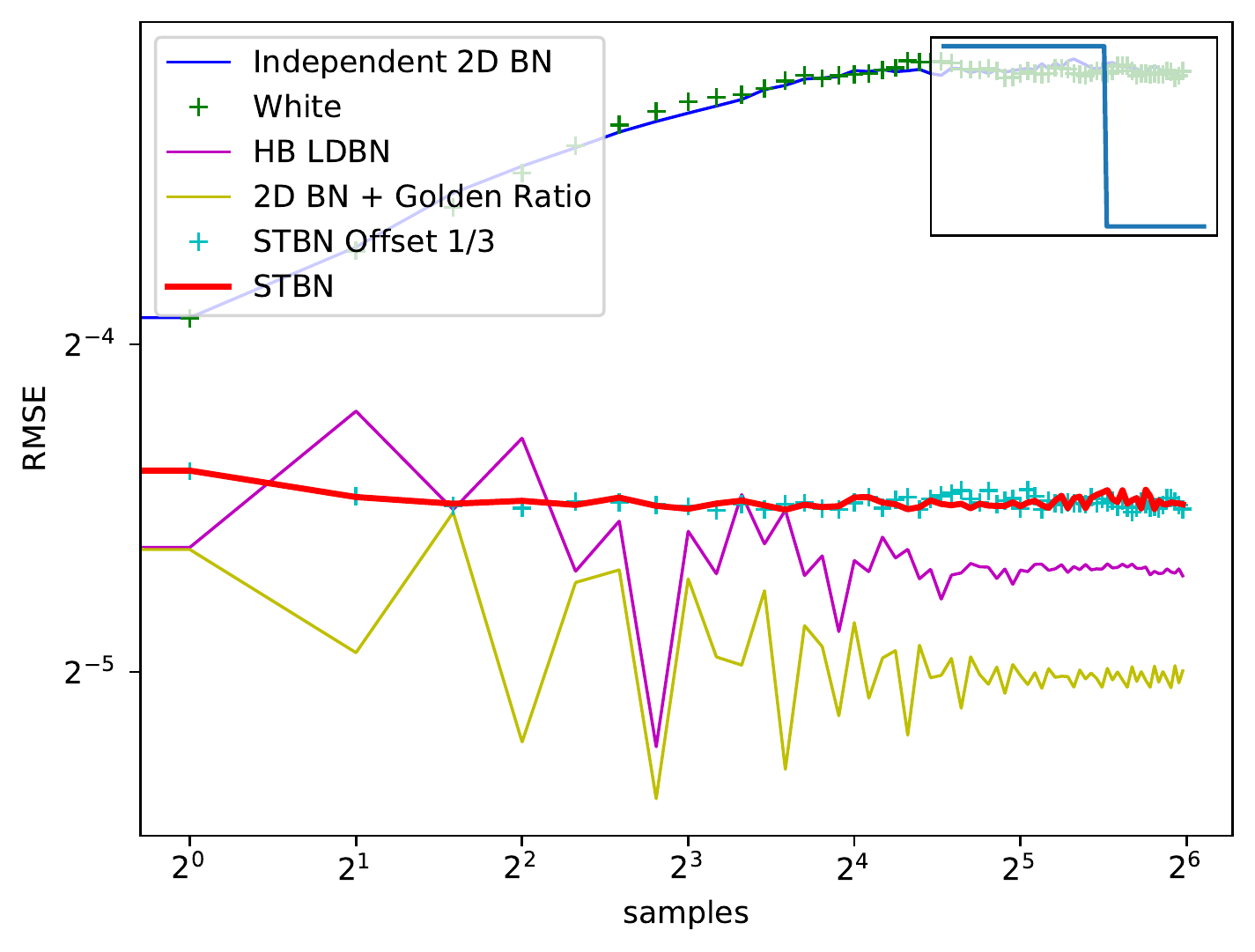} &
		\includegraphics[width=\imgW]{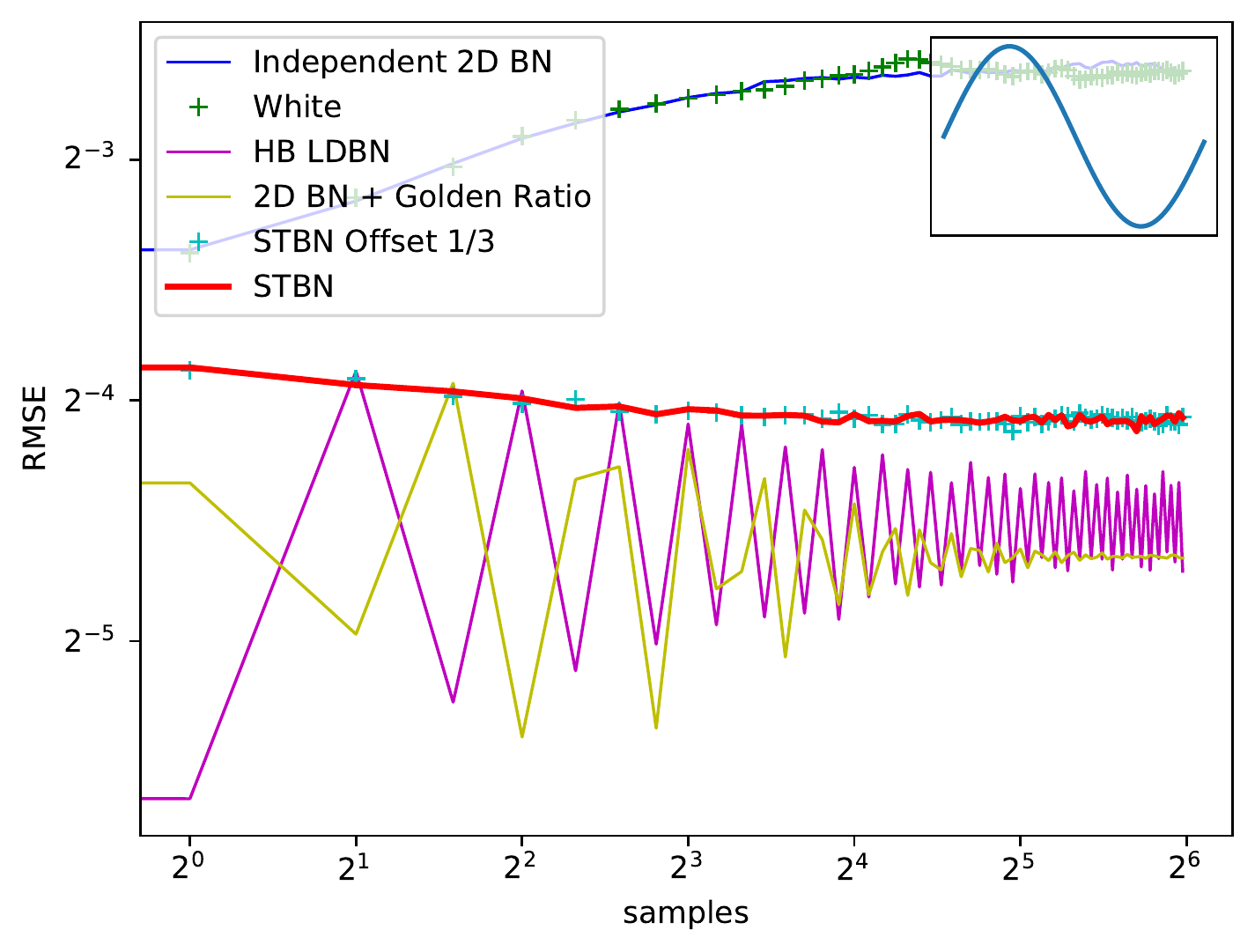} \\
		\end{tabular}
	}
	\vspace*{-.2in}
    \caption{RMSE for sampling ramp, step, and sine functions with 64 samples, combined using exponential
	moving average with $\alpha=0.1$ to simulate TAA without reprojection, motion vectors, or history rejection.
	As in Fig.~\ref{fig_3dmask_convergence_mc}, independent slices of 2D blue noise show that they
	are white noise over time, STBN shows its toroidal progressiveness, and the low discrepancy samples
	show good but erratic convergence. The poor sampling of white noise over time causes EMA per
	pixel to diverge, causing increased variance (RMSE). STBN and the low discrepancy sampling
	converge, with STBN doing so much more smoothly.
	}
	\label{fig_3dmask_convergence_taa}
\end{figure*}
Real-time temporal filtering is commonly done through an exponential moving average (EMA)~\cite{Yang2020}, rather than the more traditional Monte Carlo integration used in offline rendering. 
Therefore, we consider both EMA as well as Monte Carlo integration when analyzing results here and in the rest of the paper.

Our spatiotemporal blue noise exhibits a 1D blue noise distribution over time,
which can be seen in the upper right image (DFT(XZ)) in Figure~\ref{fig_3dmask_dfts}.
While in theory 1D blue noise distributions are not an ideal 1D sequence for Monte Carlo integration, our results show that these 1D blue 
noise sequences perform competitively in terms of convergence to low discrepancy sequences (LDS), especially at low sample counts.
As shown in Fig.~\ref{fig_3dmask_convergence_mc}, when integrating a variety of 1D functions, our 1D blue noise sequence 
converges much better than the white noise sequence produced by independent 2D blue noise texture sequences. 
When compared to the LDS optimized blue noise by Heitz and Belcour~\shortcite{Heitz2019} (HB LDBN) as well as to the golden ratio (GR) LDS, 
our 1D blue noise sequence converges competitively, although on average our results are further from the ground truth.
However, our 1D blue noise sequence converges much more stably. The convergence instability exhibited by both HB LDBN and GR LDS 
is likely due to the strobing patterns present in these sequences over time (seen as horizontal lines in the DFT(XY) in Fig.~\ref{fig_3dmask_dfts}). 


Under EMA, the results in Fig.~\ref{fig_3dmask_convergence_taa} show that our spatiotemporal blue noise can converge closer to the ground truth than the golden ratio LDS in some cases, but not all.
Compared to the Heitz and Belcour LDS, spatiotemporal blue noise is competitive, but from our initial tests a blue noise sequence incurs more error on average.
However, our spatiotemporal blue noise is much more stable over time than either of these LDS, demonstrating improved temporal coherency.

EMA can be seen as a random walk that loosely follows the values of the given samples. 
As a result, EMA is much more sensitive to clumps in the signal, which can cause low frequency strobing of pixels and increased variance.
The even sampling nature of both 1D blue noise sequences and low discrepancy sequences results in a more representative sampling of the underlying function, 
and causes the resulting EMA to converge much closer to the correct value.
Conversely, the uneven sampling of white noise causes the EMA to divergence away from the ground truth.
This is shown in Fig.~\ref{fig_3dmask_convergence_taa}, where independent blue noise diverges away from the ground truth 
due the individual textures exhibiting a white noise frequency over time, while the other sequences converge to lower error values.


\subsection{Special Qualities for Temporal Antialiasing}
\label{sec_qualities_taa}

Temporal antialiasing works by using an exponential moving average (EMA)
for each pixel, where pixels are correlated from frame to frame to account for camera and object movement.
If a pixel determines that its history is invalid due to events like disocclusions, it will restart the
integration. This is a problem if using a global progressive sequence because pixels that restart will
be using the sequence from the middle instead of the beginning, defeating the benefits of the progressive
sampling. To counter this, sequences are restarted fairly often, limiting the number of effective
samples a pixel can integrate, and thus limiting image quality. This is most commonly seen with the
sequences used for sub pixel camera jitter~\cite{Yang2020}.

Storing the index that each pixel is on individually is impractical for the storage costs required
and impossible in the case of sub pixel camera jitter. Furthermore, history rejection is not
a discrete event, but is a softer restart where the history is clamped to a color cube of values
seen in a small neighborhood for each pixel. A heuristic would be needed to decide when to actually
reset an index.

Spatiotemporal blue noise avoids this problem by being toroidally progressive on the time axis, which
also means there is no discontinuity when reaching the end of the sequence and starting over.
Hence, pixels can reject history on their own timelines and instantly start sampling with a good sequence,
which opens the door to longer sampling sequences under TAA, allowing higher image quality to be achieved.
The toroidal progressiveness can be seen in Fig.~\ref{fig_3dmask_convergence_mc} and Fig.~\ref{fig_3dmask_convergence_taa}
where ``STBN Offset 1/3'' is the same sequence as ``STBN'' but is offset 1/3 of the way through its
sequence (21 indices), and yet shows equivalent convergence behavior.

\begin{figure}[tb]
	\centering
	\setlength{\fboxsep}{0pt}%
	\setlength{\fboxrule}{0.0mm}%
	\setlength{\tabcolsep}{1.5pt}%
	\renewcommand{\arraystretch}{0.95}
	\newcommand{\imgW}{0.24\columnwidth}
	{\scriptsize 
		\begin{tabular}{cccc}
			BNOT 1/16 & Ours 1/16 & Ours 1/8 & Ours 1/32 \\
			\includegraphics[width=\imgW]{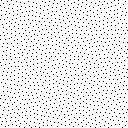} &
			\includegraphics[width=\imgW]{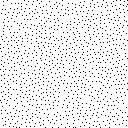} &
			\includegraphics[width=\imgW]{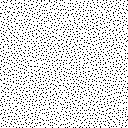} &
			\includegraphics[width=\imgW]{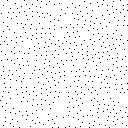} \\
		\end{tabular}
	}
	\caption{Comparing 1024 BNOT samples with a slice of a $128\times 128\times 10$ spatiotemporal blue noise mask (ours) thresholded
	to different levels. BNOT is much higher quality over space, but has a fixed density and gives no
	treatment to the time axis, necessitating independent sample sets to be white noise over time.}
	\label{fig_points_vsbnot_space}
\end{figure}

\begin{figure}[tb]
	\centering
	\setlength{\fboxsep}{0pt}%
	\setlength{\fboxrule}{0.0mm}%
	\setlength{\tabcolsep}{1.5pt}%
	\renewcommand{\arraystretch}{0.95}
	\newcommand{\imgW}{0.24\columnwidth}
	{\scriptsize 
		\begin{tabular}{cccc}
			XY & DFT(XY) & YZ & DFT(YZ) \\
			\includegraphics[width=\imgW]{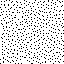} &
			\includegraphics[width=\imgW]{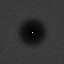} &
			\includegraphics[width=\imgW]{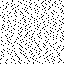} &
			\includegraphics[width=\imgW]{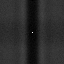} \\
		\end{tabular}
	}
	\caption{2D projections of our $64\times 64\times 64$ spatiotemporal blue noise (STBN) masks thresholded to point sets, which
	keep their desired spectra. DFTs are averaged over 64 slices.}
	\label{fig_points_vsbnot_time}
\end{figure}

\begin{figure*}[tb]
	\centering
	\setlength{\fboxsep}{0pt}%
	\setlength{\fboxrule}{0.0mm}%
	\setlength{\tabcolsep}{1.5pt}%
	\renewcommand{\arraystretch}{0.95}
	\newcommand{\imgW}{0.17\textwidth}
	{\scriptsize 
		\begin{tabular}{ccccc}
			Importance Map & White noise & 2DBN &
			STBN (ours) &
			\multirow{2}{*}{\includegraphics[width=0.27\textwidth]{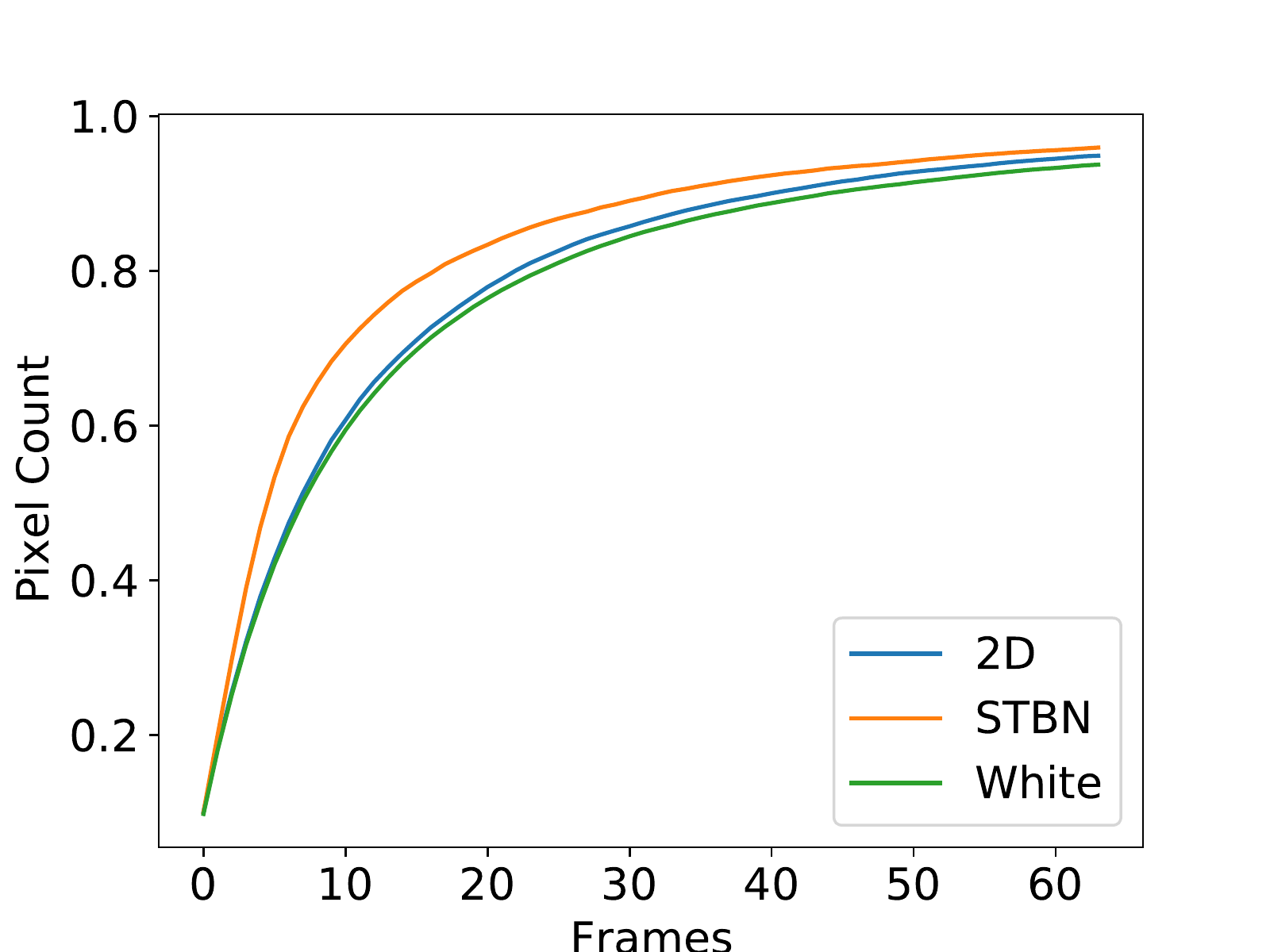}}
			\\
			\includegraphics[width=\imgW]{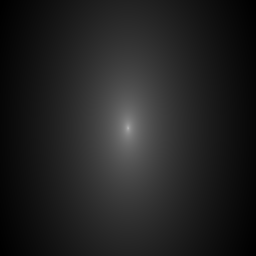} &
			\includegraphics[width=\imgW]{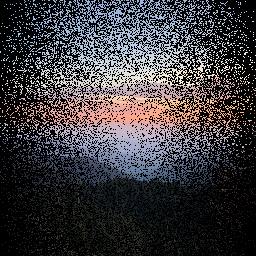} &
			\includegraphics[width=\imgW]{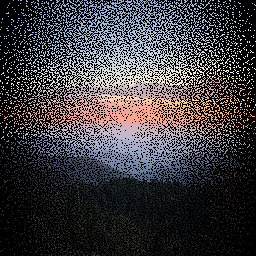} &
			\includegraphics[width=\imgW]{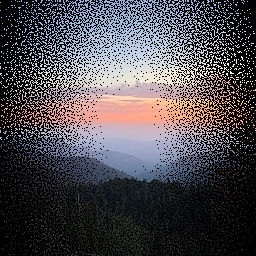} &
		\end{tabular}
	}
	\caption{To the left, we show an importance map, which was used to generate the next three images.
	These were generated by simulating adaptive sampling using the importance map to guide where the samples
	should be over five frames. The images were generated in such a way that the more pixels that are lit, the more unique
	pixels have been sampled. STBN provides the best results here, since it has substantially more lit pixels.
	Right: a diagram showing the percentage of unique pixels sampled over time as a function of frame number.
	White noise can have redundant sampled pixels each frame, and over time, 2D blue
	noise reduces redundant pixels over space, and STBN (ours) reduces them over
	time as well.
	}
	\label{fig_points_uniquepixels_render}
\end{figure*}

\subsection{Spatiotemporal Point Sets}
\label{sec_pointsets}

Since our spatiotemporal blue noise masks are created using the void and cluster algorithm, they 
have the property that
if you threshold the mask values to some percentage, the same percentage of the pixels will survive.
Thresholding a blue noise mask produces a \textit{sample point set}.
Our thresholded masks produce blue noise sample patterns over both space and time.
While there are better algorithms for generating blue noise sample patterns, such as blue noise
through optimal transport (BNOT)~\cite{deGoes2012}, they do not handle the time axis. To the best of our
knowledge, our algorithm is the first method for making spatiotemporal blue noise sample patterns.

Figure~\ref{fig_points_vsbnot_space} shows how thresholded masks are not as high quality spatially as BNOT,
but they are able to make point sets of any density.  Figure~\ref{fig_points_vsbnot_time} shows how the
thresholded point sets keep their desired frequency spectra over axis groups.

An example usage case is doing sparse ray tracing, using an importance map to 
decide which regions should have more samples. For each pixel $\mathbf{p}$, we shoot a ray if and
only if the importance map value $f(\mathbf p) \in [0,1]$ is greater than the mask value $g(\mathbf p) \in [0,1]$.
Using white noise over space and time, the spatial pattern will have redundant samples and large holes, while also having duplicated samples
over time.  Using spatial blue noise makes the sampling more even over space, but still has the
problems of white noise over time.  Using spatiotemporal blue noise means that samples are evenly
distributed over both space and time. This is shown in Fig.~\ref{fig_points_uniquepixels_render}.

Two fundamental direct usage cases of blue noise masks, generated by the void and cluster algorithm,
are stippling and dithering.
Whenever you are stochastically turning a continuous probability into a discrete 0 or a 1,
you are using the stippling property.  We will cover these two direct usage cases in the next section,
where stochastic transparency, which is a type of stippling, is presented in Section~\ref{sec_applications_alpha}
and spatiotemporal dithering in Section~\ref{sec_applications_dithering}.



\section{Applications and Results}
\label{sec_applications}

In this section, we demonstrate using our spatiotemporal blue noise masks in a range of techniques and
the results of using them. While there are other ways to implement the techniques shown, our niche
is low sample count and low overhead real-time rendering applications.
We compare in the context of both Monte Carlo integration, which is used when multiple samples are
taken per frame, and in the context of temporal antialiasing (TAA), or EMA when there is no
depth buffer or motion vectors.

We start with stochastic transparency because it requires only one random value per pixel, then move to
dithering, which requires three random values per pixel. We then show an example with 
volumetric rendering, and finally show ray traced ambient occlusion (AO), which needs a 2D vector per pixel. These
techniques are shown to demonstrate concepts of using spatiotemporal blue noise, which can be carried over
to other modern rendering techniques.

\subsection{Stochastic Transparency}
\label{sec_applications_alpha}

Stochastic transparency is the process of stochastically choosing whether to ignore a sample
based on a material's transparency level. This is useful in situations such as deferred lighting
where you are storing information about how to shade a pixel, instead of the shaded result itself,
and it is impractical to store multiple layers to later calculate proper transparency.

Sophisticated algorithms have been developed by Enderton et al.~\shortcite{Enderton2011},
Wyman and McGuire~\shortcite{Wyman2017}, and are also discussed by Wyman~\shortcite{Wyman2016}, but the core idea
of stochastically accepting or rejecting a pixel remains the same. Our algorithm described here is
aimed at low computational costs (a single texture read and comparison), giving blue noise distributed
error in screen space as 2D blue noise does, but also converging faster.
We show renderings of stochastic alpha using various types of spatiotemporal noise under both Monte
Carlo integration and EMA in Fig.~\ref{fig_stochastic_alpha_renders}. 

\setlength{\fboxsep}{0pt}
\setlength{\fboxrule}{1pt}
\newbox{\bigpicturebox}
\begin{figure*}[tb]
	\centering
	\setlength{\tabcolsep}{1.5pt}%
	\newcommand{\imgW}{0.08\textwidth}
	{\scriptsize
	\begin{tabular}{ccccccccc}
	STBN (Ours) 1 Frame & & Frame 13 & DFT(XY) & ZY & DFT(ZY) & MC 4~Frames & EMA 64~Frames \\
	\raisebox{12.8mm}{\multirow{4}{*}{
		\setlength{\fboxrule}{0pt}
		\fbox{\adjustbox{trim={0} {0} {0cm} {0},clip}{
			\includegraphics[width=.445\textwidth]{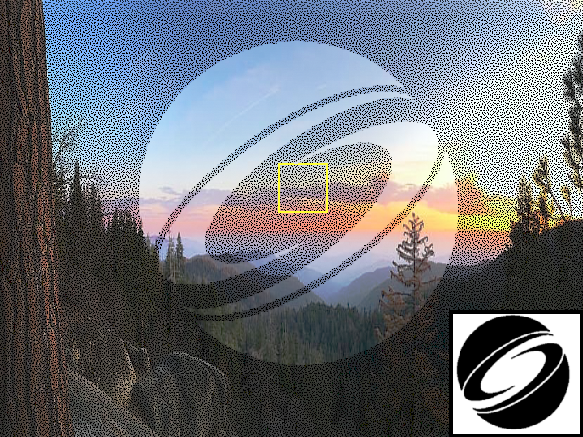}}
		}}
	}
	\hspace{1mm}
	&
	\rotatebox{90}{White}&
	\includegraphics[width=\imgW]{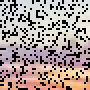} &
	\includegraphics[width=\imgW]{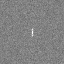} &
	\includegraphics[width=\imgW]{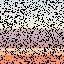} &
	\includegraphics[width=\imgW]{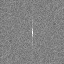} &
	\includegraphics[width=\imgW]{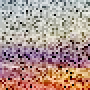} &
	\includegraphics[width=\imgW]{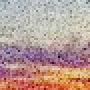} &
	\\
	&
	\rotatebox{90}{2DBN}&
	\includegraphics[width=\imgW]{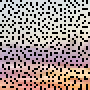} &
	\includegraphics[width=\imgW]{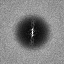} &
	\includegraphics[width=\imgW]{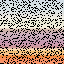} &
	\includegraphics[width=\imgW]{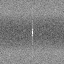} &
	\includegraphics[width=\imgW]{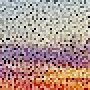} &
	\includegraphics[width=\imgW]{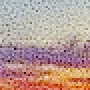} &
	\\
	&
	\rotatebox{90}{GR}&
	\includegraphics[width=\imgW]{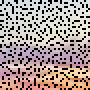} &
	\includegraphics[width=\imgW]{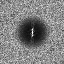} &
	\includegraphics[width=\imgW]{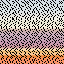} &
	\includegraphics[width=\imgW]{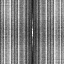} &
	\includegraphics[width=\imgW]{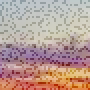} &
	\includegraphics[width=\imgW]{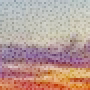} &
	\\
	&
	\rotatebox{90}{STBN (ours)}&
	\includegraphics[width=\imgW]{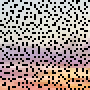} &
	\includegraphics[width=\imgW]{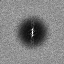} &
	\includegraphics[width=\imgW]{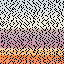} &
	\includegraphics[width=\imgW]{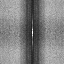} &
	\includegraphics[width=\imgW]{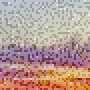} &
	\includegraphics[width=\imgW]{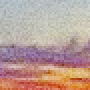} &
	\end{tabular}
	}
	\vspace*{-.2in}
	\caption{Stochastic transparency with $\alpha=0.9$ using various noise types and integration schemes.
	At 4 frames of Monte Carlo integration, our STBN provides better quality than 2D blue noise, and is competitive
	with golden ratio (GR) animated blue noise. Under 64 frames of exponential moving average (EMA), our STBN provides
	better quality than golden ratio noise. Columns 2 and 4 are averaged DFTs to show expected spectra.}
	\label{fig_stochastic_alpha_renders}
\end{figure*}

\setlength{\fboxsep}{0pt}
\setlength{\fboxrule}{1pt}
\newbox{\bigpicturebox}
\begin{figure*}[tb]
	\centering
	\setlength{\tabcolsep}{1.5pt}%
	\newcommand{\imgW}{0.08\textwidth}
	{\scriptsize
	\begin{tabular}{ccccccccc}
	STBN (Ours) 1 Frame & & 1~Frame & DFT(XY) & ZY & DFT(ZY) & MC 4~Frames & EMA Frame 64 \\
	\raisebox{12.8mm}{\multirow{4}{*}{
		\setlength{\fboxrule}{0pt}
		\fbox{\adjustbox{trim={0} {0} {0cm} {0},clip}{
			\includegraphics[width=.445\textwidth]{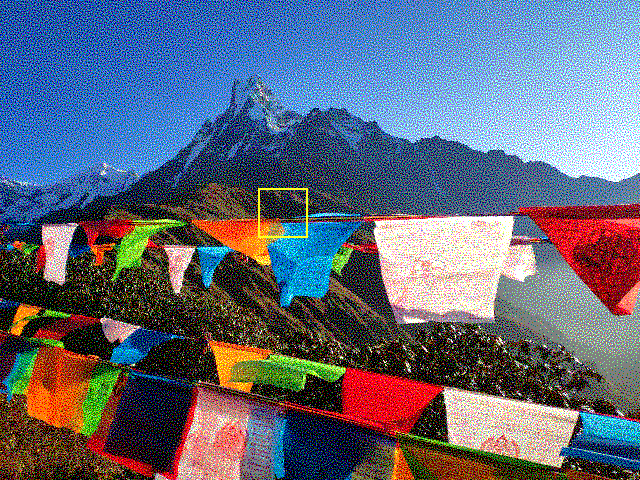}}
		}}
	}
	\hspace{1mm}
	&
	\rotatebox{90}{White}&
	\includegraphics[width=\imgW]{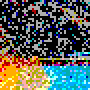} &
	\includegraphics[width=\imgW]{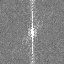} &
	\includegraphics[width=\imgW]{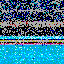} &
	\includegraphics[width=\imgW]{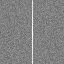} &
	\includegraphics[width=\imgW]{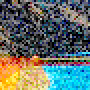} &
	\includegraphics[width=\imgW]{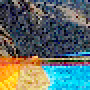} &
	\\
	&
	\rotatebox{90}{2DBN}&
	\includegraphics[width=\imgW]{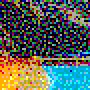} &
	\includegraphics[width=\imgW]{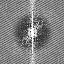} &
	\includegraphics[width=\imgW]{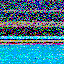} &
	\includegraphics[width=\imgW]{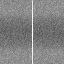} &
	\includegraphics[width=\imgW]{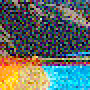} &
	\includegraphics[width=\imgW]{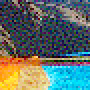} &
	\\
	&
	\rotatebox{90}{GR}&
	\includegraphics[width=\imgW]{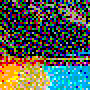} &
	\includegraphics[width=\imgW]{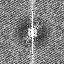} &
	\includegraphics[width=\imgW]{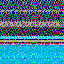} &
	\includegraphics[width=\imgW]{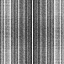} &
	\includegraphics[width=\imgW]{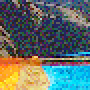} &
	\includegraphics[width=\imgW]{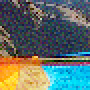} &
	\\
	&
	\rotatebox{90}{STBN (ours)}&
	\includegraphics[width=\imgW]{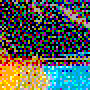} &
	\includegraphics[width=\imgW]{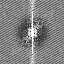} &
	\includegraphics[width=\imgW]{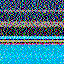} &
	\includegraphics[width=\imgW]{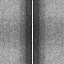} &
	\includegraphics[width=\imgW]{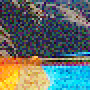} &
	\includegraphics[width=\imgW]{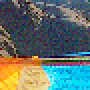} &
	\end{tabular}
	}
	\vspace*{-.2in}
	\caption{Dithering to 1 bit per color channel using various noise types and integration schemes.
	Golden ratio animated blue noise and our spatiotemporal blue noise generate images with better quality than the other types of
	noise, and look
	comparable under Monte Carlo integration. Our STBN generates better image quality than golden ratio under 64 frames
	of TAA, however.}
	\label{fig_dither_renders}
\end{figure*}

\begin{figure}[tb]
	\centering
	\setlength{\fboxsep}{0pt}%
	\setlength{\fboxrule}{0.0mm}%
	\setlength{\tabcolsep}{1.5pt}%
	\renewcommand{\arraystretch}{0.95}
	\newcommand{\imgW}{0.49\columnwidth}
	{\scriptsize 
		\begin{tabular}{cc}
			Monte Carlo Integration & EMA \\
			\includegraphics[width=\imgW]{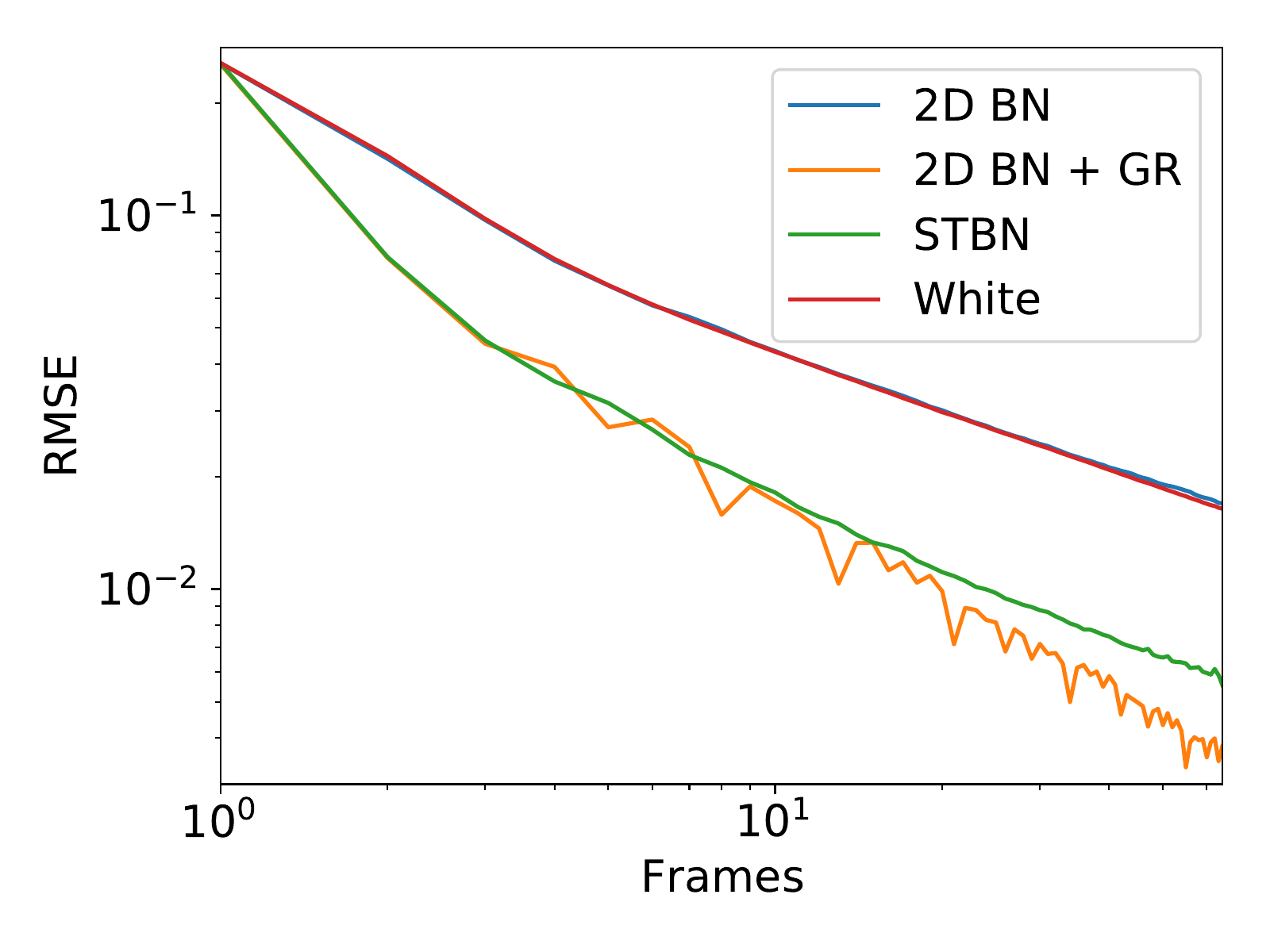} &
			\includegraphics[width=\imgW]{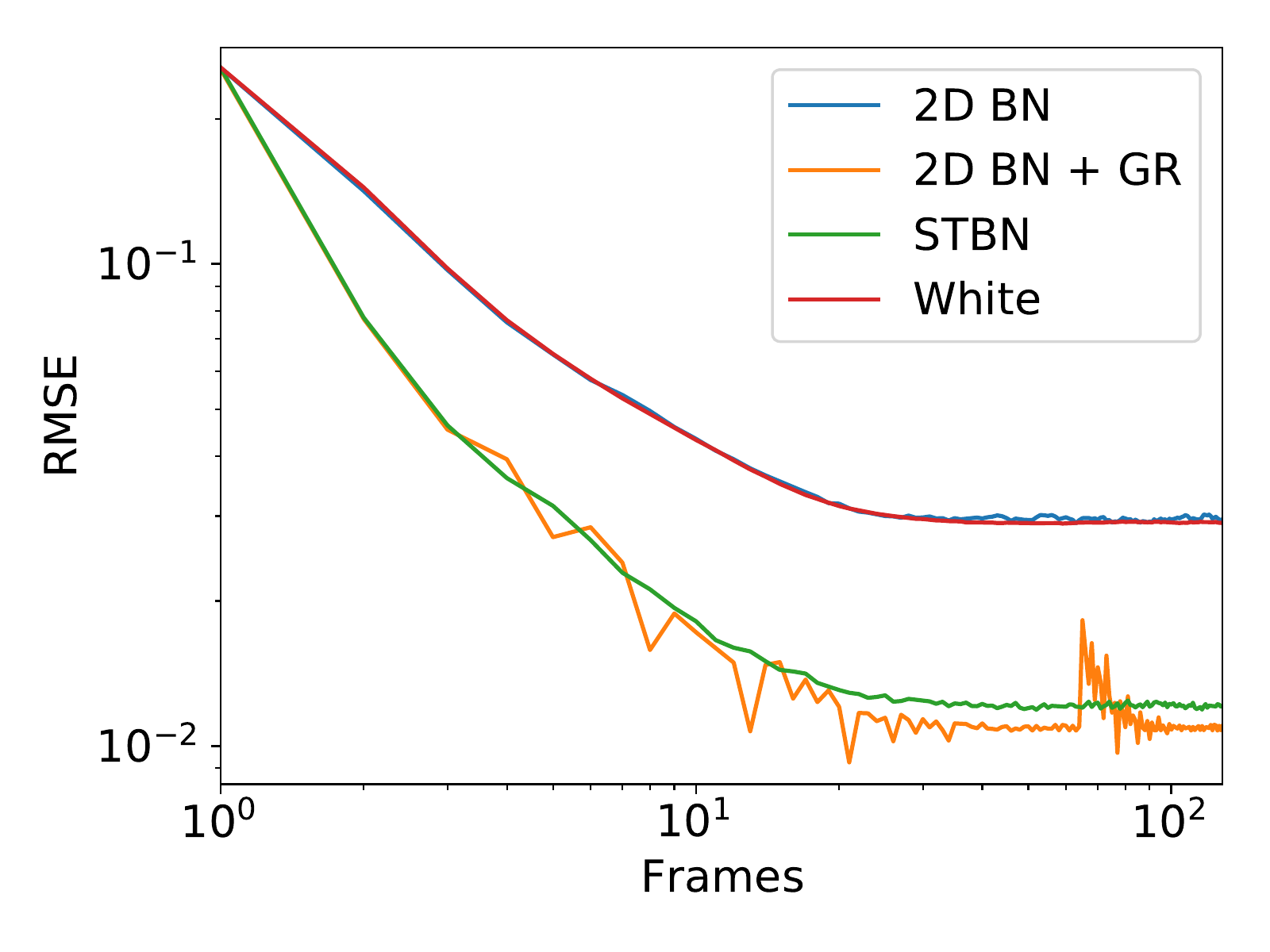} \\
		\end{tabular}
	}
	\vspace*{-.2in}
	\caption{Convergence rates in stochastic transparency of various types of noise.  Golden ratio (GR) animated
	blue noise converges marginally faster than our spatiotemporal blue noise (STBN), but damages frequencies spatially,
	and has an error spike under EMA when the sequence needs to restart to avoid numerical issues.}
	\label{fig_stochastic_alpha_convergence}
\end{figure}
\begin{figure}[tb]
	\centering
	\includegraphics[width=0.8\columnwidth]{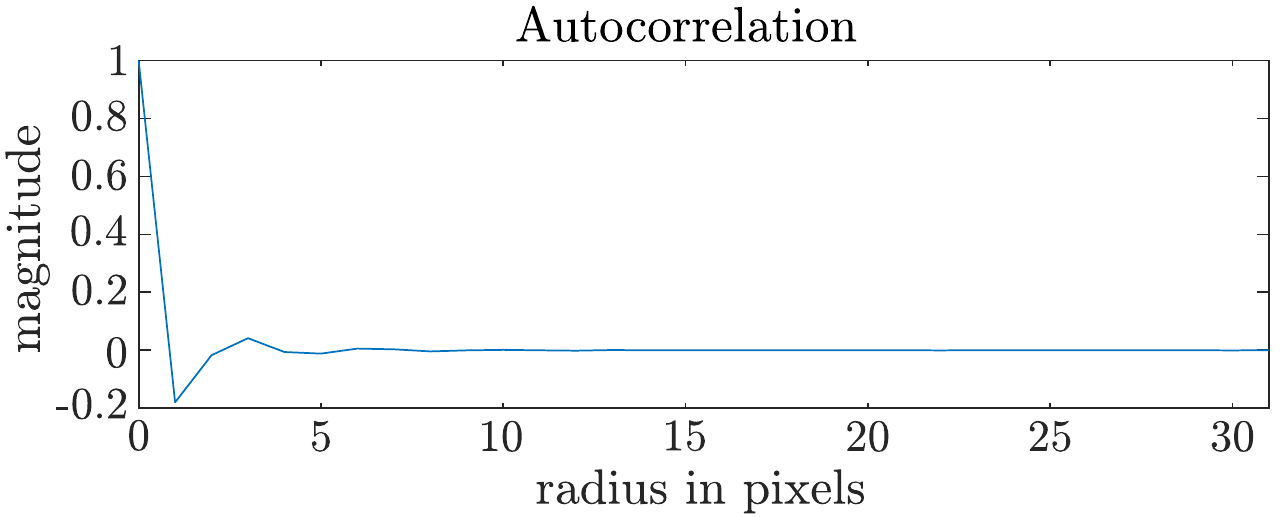}
	\vspace*{-2mm}
	\caption{For this diagram, we computed the autocorrelation of 64 blue noise textures of 
	resolution $64\times 64$, which resulted in a circularly symmetric image. We show a radial plot of the average of those.
	In blue noise textures, neighbors have very different values, which causes a rippling of correlation
	and anti-correlation for small values of the radius, but rapidly decays to zero, i.e., to decorrelated values.
	}
	\label{fig_bn_autocorrelation}
	\vspace*{-3mm}
\end{figure}

When looking at frame 13 in isolation, independent 2D blue noise and spatiotemporal blue noise have the same
quality which is correct and shows that our noise is as good for each slice in time. Under
4 frames of Monte Carlo or 64 frames of EMA, our spatiotemporal blue noise shows much better
convergence than independent blue noise.  Our method is competitive with golden ratio animated blue noise for
4 frames of Monte Carlo, but is superior at 64 frames of EMA where the golden ratio sequence restarts.
Under motion, the golden ratio shows high-frequency strobing as shown in the supplemental material.
Convergence graphs are shown in Fig.~\ref{fig_stochastic_alpha_convergence}, which reveal the same
story as what is seen visually.

\subsection{Dithering}
\label{sec_applications_dithering}

Dithering is the process of adding a small random value before quantization to turn quantization
artifacts (banding) into noise instead. This allows less memory to be used while preserving image
quality.  Dithering rounds the quantized value up or down randomly, with probability to round down
being higher as the value gets closer to the lower boundary of quantization.
White noise is not often prefered in dithering, but Bayer~\shortcite{Bayer1973} and the less real-time friendly techniques,
such as Floyd-Steinberg error diffusion~\shortcite{Steinberg1976}, do not consider the time axis. For
a more in depth read about dithering, consult Christou's thesis~\shortcite{Christou2008}.

The rendered results in Fig.~\ref{fig_dither_renders} reveal that our noise deliver approximately 
the same quality as
golden ratio animated noise and much better than the other types of noise. Our noise again provides
better image quality at 64 frames of EMA though. As before, the golden ratio noise has high-frequency
strobing as can be seen in the supplemental material. Convergence graphs are nearly identical to
Fig.~\ref{fig_stochastic_alpha_convergence}, and so are omitted.

\setlength{\fboxsep}{0pt}
\setlength{\fboxrule}{1pt}
\newbox{\bigpicturebox}
\begin{figure*}[tb]
	\centering
	\setlength{\tabcolsep}{1.5pt}%
	\newcommand{\imgW}{0.08\textwidth}
	{\scriptsize
	\begin{tabular}{ccccccccc}
	STBN (Ours) 1 Frame & & 1~Frame & DFT(XY) & ZY & DFT(ZY) & MC 4~Frames & EMA Frame 64 \\
	\raisebox{12.8mm}{\multirow{4}{*}{
		\setlength{\fboxrule}{0pt}
		\fbox{\adjustbox{trim={0cm} {0} {0cm} {0},clip}{
			\includegraphics[width=.445\textwidth]{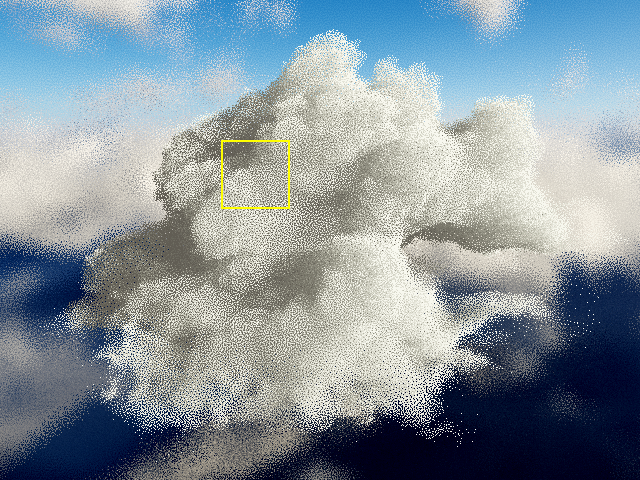}}
		}}
	}
	\hspace{1mm}
	&
	\rotatebox{90}{White}&
	\includegraphics[width=\imgW]{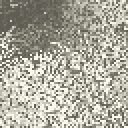} &
	\includegraphics[width=\imgW]{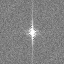} &
	\includegraphics[width=\imgW]{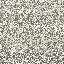} &
	\includegraphics[width=\imgW]{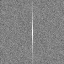} &
	\includegraphics[width=\imgW]{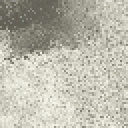} &
	\includegraphics[width=\imgW]{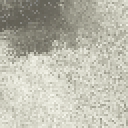} &
	\\
	&
	\rotatebox{90}{2DBN}&
	\includegraphics[width=\imgW]{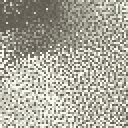} &
	\includegraphics[width=\imgW]{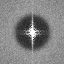} &
	\includegraphics[width=\imgW]{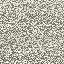} &
	\includegraphics[width=\imgW]{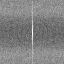} &
	\includegraphics[width=\imgW]{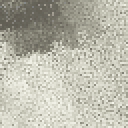} &
	\includegraphics[width=\imgW]{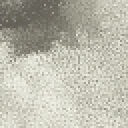} &
	\\
	&
	\rotatebox{90}{GR}&
	\includegraphics[width=\imgW]{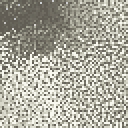} &
	\includegraphics[width=\imgW]{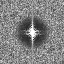} &
	\includegraphics[width=\imgW]{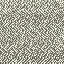} &
	\includegraphics[width=\imgW]{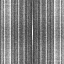} &
	\includegraphics[width=\imgW]{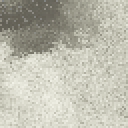} &
	\includegraphics[width=\imgW]{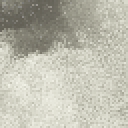} &
	\\
	&
	\rotatebox{90}{STBN (ours)}&
	\includegraphics[width=\imgW]{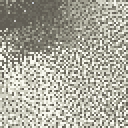} &
	\includegraphics[width=\imgW]{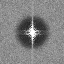} &
	\includegraphics[width=\imgW]{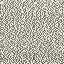} &
	\includegraphics[width=\imgW]{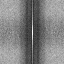} &
	\includegraphics[width=\imgW]{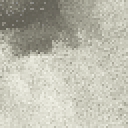} &
	\includegraphics[width=\imgW]{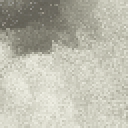} &
	\\
	\vspace*{-18.25mm}	
	\\					
	\hspace*{63.25mm}	
	\fbox{\includegraphics[width=\imgW]{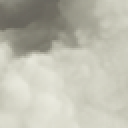}} 
	\end{tabular}
	}
	\caption{Single scattering volume rendering using various noise types and integration schemes.
	Our spatiotemporal noise and golden ratio animated blue noise are comparable, but we have slightly
	better convergence and are more temporally stable under EMA as can be seen in the
	supplemental materials. The ground truth is shown in the inset in the left image (lower right corner). 
	}
	\label{fig_volume_renders}
\end{figure*}

Dithering an RGB image requires three random values per pixel, but our spatiotemporal blue noise masks
only provide a single value per pixel. One could use three independent spatiotemporal blue noise masks,
but that could be undesirable due to increased memory usage.  
Autocorrelation can be used to find repeating patterns within a signal and
Fig.~\ref{fig_bn_autocorrelation} shows that correlation only exists over short distances in blue noise textures.
We exploit this to our advantage and use the R2 low discrepancy sequence~\cite{Roberts2018} to get three
nearly maximally spaced offsets to read the texture at to get three fairly independent spatiotemporal
blue noise values per pixel. This method can be generalized to give $N$ fairly independent spatiotemporal
blue noise values when needed.


\subsection{Single Scattering Media} 
\label{sec_applications_fog}

Next, we present an algorithm for rendering single scattering heterogeneous participating media 
with very low sample counts, similar to the airlight model by Sun et al.~\shortcite{sun2005} 
(but computed numerically for heterogenous volumes). 
This is a different type of algorithm than stochastic transparency or dithering 
because it shows how blue noise masks can be applied to a more general rendering problem.
While there are much more sophisticated algorithms for rendering participating media, our method is
simple, efficient, gives good results at very low sample counts and works with either rasterization
or ray tracing pipelines.

First, a camera ray $\mathbf o + t\vec \omega_o$ is cast through the bounds of a heterogeneous volume, where an enter distance $t_\mathrm{min}$ and 
exit distance $t_\mathrm{max}$ through those bounds are recorded. From here, we use stochastic ray marching 
to march through the volume at $n$ evenly spaced locations, where the space between each
sample is $\frac{d}{n}$ units and $d = t_\mathrm{max} - t_\mathrm{min}$.
The location $\mathbf p_s$ of a sample $s \in \mathbf{Z} [0, n-1]$ is then calculated as
\begin{equation}
	\mathbf p_s = \mathbf o +  \left(s \frac{d}{n}+t_\mathrm{min}\right)\vec \omega_o.
\end{equation}
At each sample point $\mathbf p_s$, a volumetric density field $F$ is sampled to get a density $f_s$. This density is assumed
to represent the density for an entire step length of distance.
We accumulate this density and proceed to the next sample point until the following criterion is met:
\begin{equation}
	\sum_s{f_s} \geq -\ln(1 - \xi),
\end{equation}
where $\xi$ is a drawn random number. Once this condition is met, the point on the ray $\mathbf p_s$ will be approximately located
at the sampled free-flight distance, or the distance at which the ray collides with a particle in the media. For 
more information, see the course by Nov\'ak et al.~\shortcite{novak2018}.

At this point, a second ray originating at this collision point is traced toward a directional light source.
Again, we march the ray through the volume. This second ray composites volumetric samples from front to back
until the ray exits the volume. This composited value is then used to represent transmittance 
along the ray.
To shade our collision point, we multiply the light intensity by the transmittance 
of light, and multiply by the albedo 
of the volume at the sampled collision location.

\begin{figure}[tb]
	\centering
	\setlength{\fboxsep}{0pt}%
	\setlength{\fboxrule}{0.0mm}%
	\setlength{\tabcolsep}{1.5pt}%
	\renewcommand{\arraystretch}{0.95}
	\newcommand{\imgW}{0.49\columnwidth}
	{\scriptsize 
		\begin{tabular}{ccc}
			Monte Carlo Integration & EMA \\
			\includegraphics[width=\imgW]{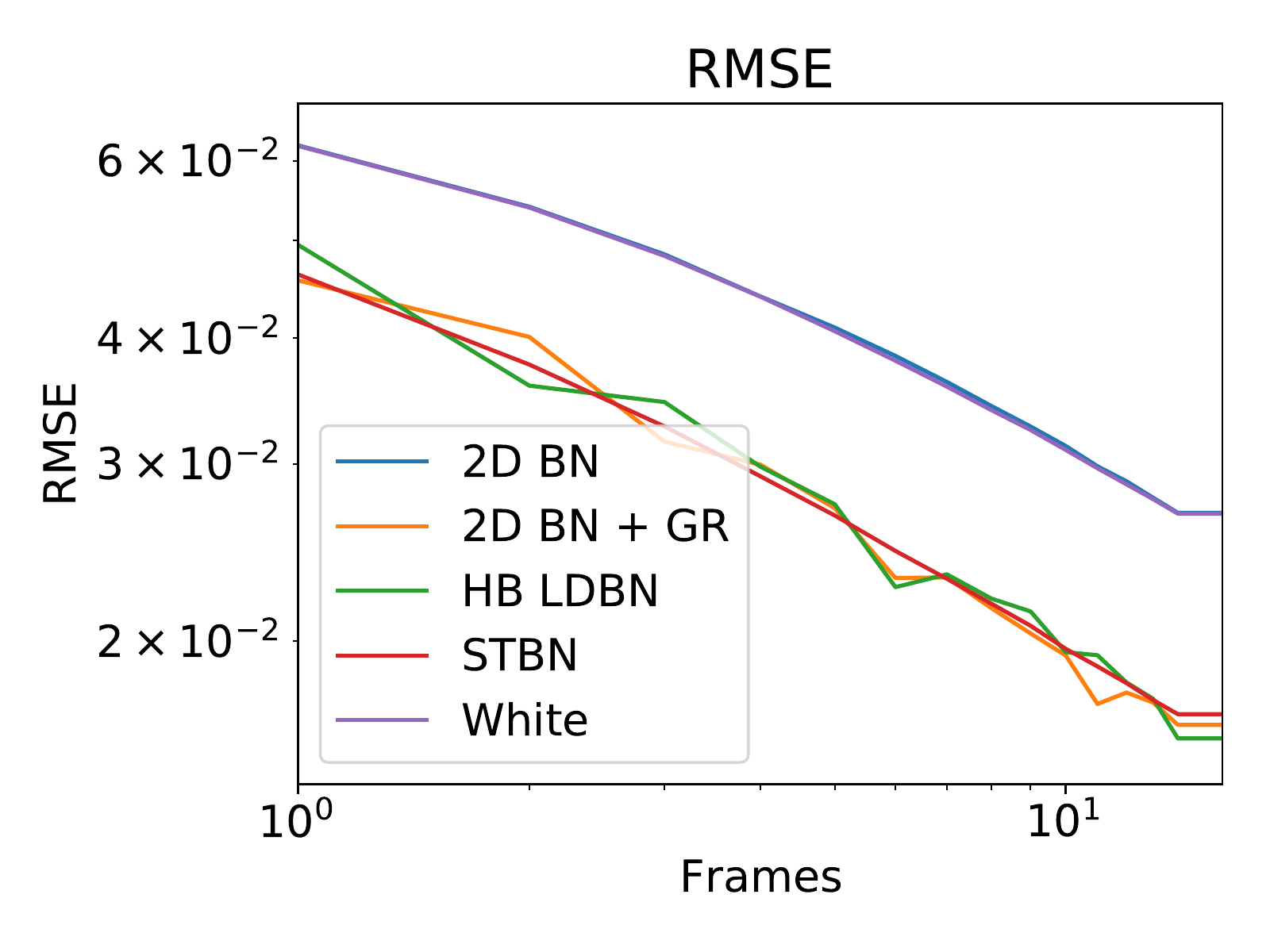} &
			\includegraphics[width=\imgW]{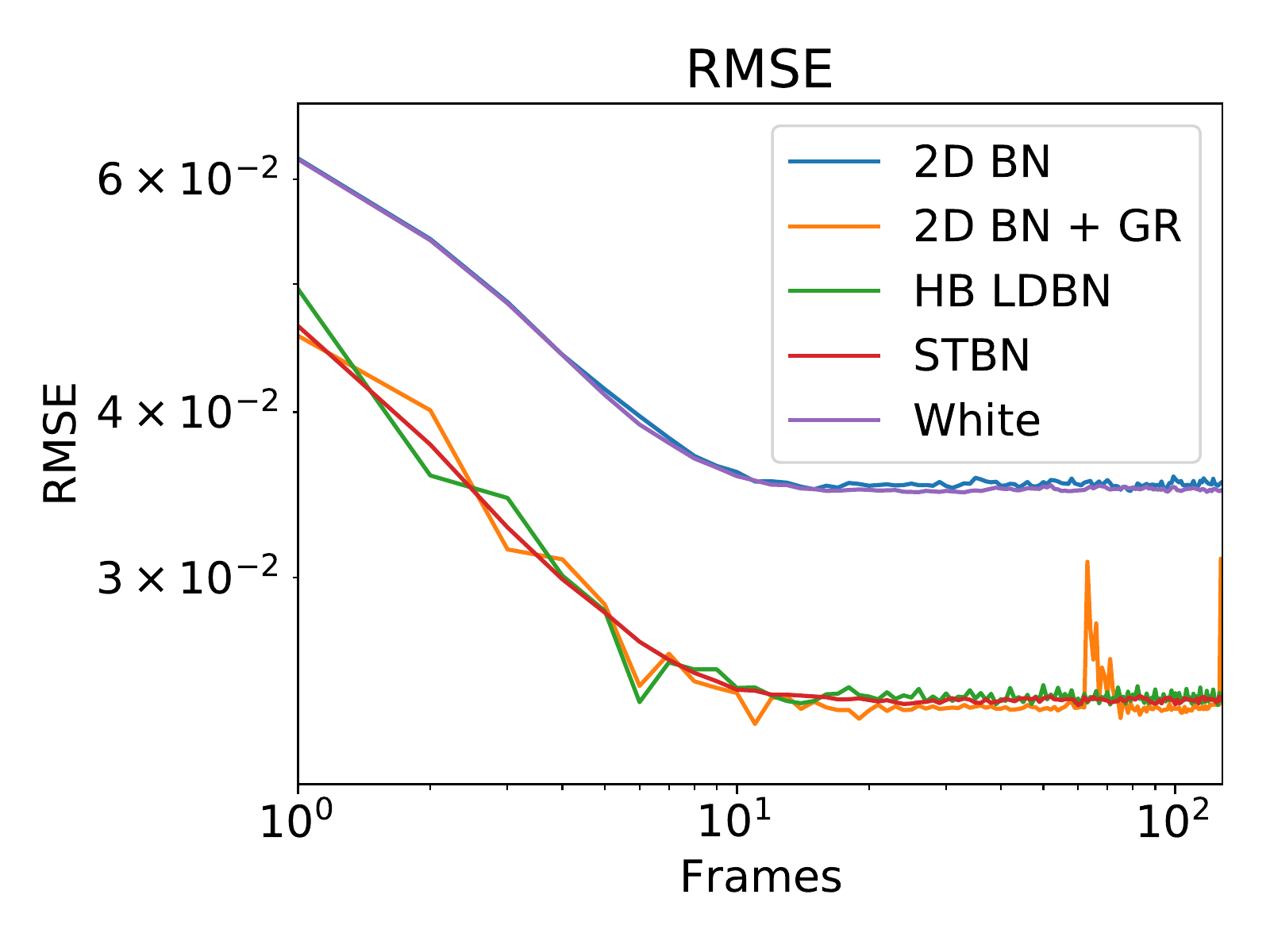} \\
		\end{tabular}
	}
	\caption{Convergence rate of single scattered volume rendering. Our spatiotemporal blue noise
	converges at essentially the same rate as the low discrepancy sequences but is much more stable,
	and does not have the seam at the sequence restart point seen in EMA.}
	\label{fig_volume_convergence}
\end{figure}

\setlength{\fboxsep}{0pt}
\setlength{\fboxrule}{1pt}
\newbox{\bigpicturebox}
\begin{figure*}[tb]
	\setlength{\tabcolsep}{1.5pt}%
	\newcommand{\imgW}{0.08\textwidth}
	{\scriptsize
	\centering
	\begin{tabular}{ccccccccc}
	STBN (Ours) 1 Frame & & 1~Frame & DFT(XY) & ZY & DFT(ZY) & MC 16~Frames & EMA Frame 64 \\
	\raisebox{12.8mm}{\multirow{4}{*}{
		\setlength{\fboxrule}{0pt}
		\frame{\setlength{\fboxsep}{0pt}\fbox{\adjustbox{trim={0} {0} {0cm} {0},clip}{\includegraphics[width=.445\textwidth]{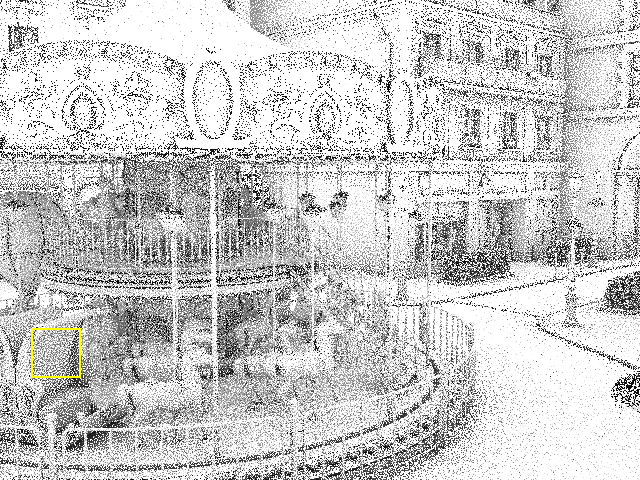}}}}}
	}
	\hspace{1mm}
	&
	\rotatebox{90}{White}&
	\includegraphics[width=\imgW]{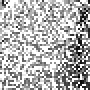} &
	\includegraphics[width=\imgW]{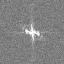} &
	\includegraphics[width=\imgW]{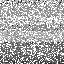} &
	\includegraphics[width=\imgW]{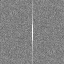} &
	\includegraphics[width=\imgW]{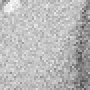} &
	\includegraphics[width=\imgW]{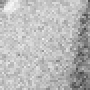} &
	\\
	&
	\rotatebox{90}{2DBN}&
	\includegraphics[width=\imgW]{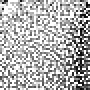} &
	\includegraphics[width=\imgW]{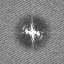} &
	\includegraphics[width=\imgW]{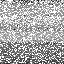} &
	\includegraphics[width=\imgW]{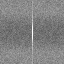} &
	\includegraphics[width=\imgW]{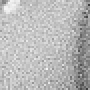} &
	\includegraphics[width=\imgW]{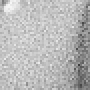} &
	\\
	&
	\rotatebox{90}{GR}&
	\includegraphics[width=\imgW]{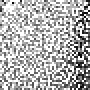} &
	\includegraphics[width=\imgW]{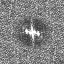} &
	\includegraphics[width=\imgW]{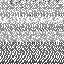} &
	\includegraphics[width=\imgW]{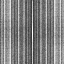} &
	\includegraphics[width=\imgW]{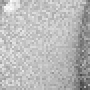} &
	\includegraphics[width=\imgW]{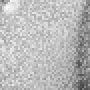} &
	\\
	&
	\rotatebox{90}{STBN (ours)}&
	\includegraphics[width=\imgW]{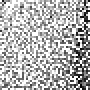} &
	\includegraphics[width=\imgW]{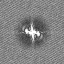} &
	\includegraphics[width=\imgW]{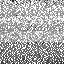} &
	\includegraphics[width=\imgW]{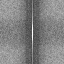} &
	\includegraphics[width=\imgW]{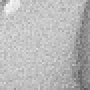} &
	\includegraphics[width=\imgW]{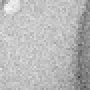} &
	\vspace*{-15.5mm}	
	\\					
	\hspace*{63.5mm}	
	\fbox{\includegraphics[width=\imgW]{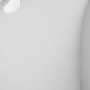}} 
	\end{tabular}	
	}
	\vspace*{-1mm}
	\caption{Ray traced ambient occlusion using various noise types and integration schemes.
	STBN uses independent spatiotemporal blue noise for the $x$ and $y$ axes.
	Golden ratio provides lower image quality here because it ends up being the same low discrepancy sequence for $x$ and $y$.
	STBN looks the most converged under both 16 frames of Monte Carlo, as well as 64 frames of EMA.
	Columns 2 and 4 are averaged to show expected spectra. 
	The ground truth is shown in the inset in the left image (lower right corner). 
	}
	\label{fig_ao_renders}
\end{figure*}

\begin{figure}[tb]
	\centering
	\setlength{\fboxsep}{0pt}%
	\setlength{\fboxrule}{0.0mm}%
	\setlength{\tabcolsep}{1.5pt}%
	\renewcommand{\arraystretch}{0.95}
	\newcommand{\imgW}{0.49\columnwidth}
	{\scriptsize 
		\begin{tabular}{ccc}
			Monte Carlo Integration & TAA \\
			\includegraphics[width=\imgW]{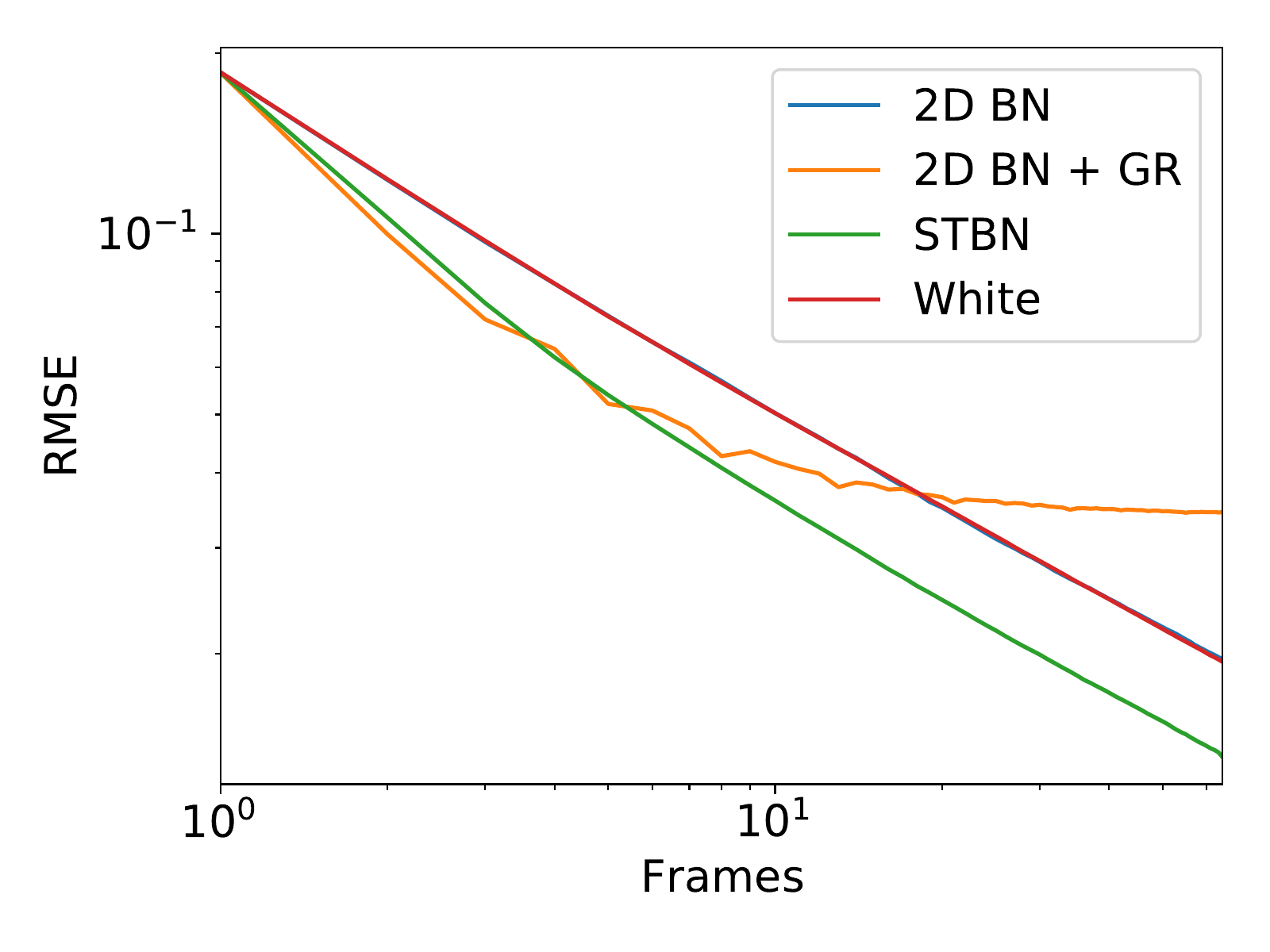} &
			\includegraphics[width=\imgW]{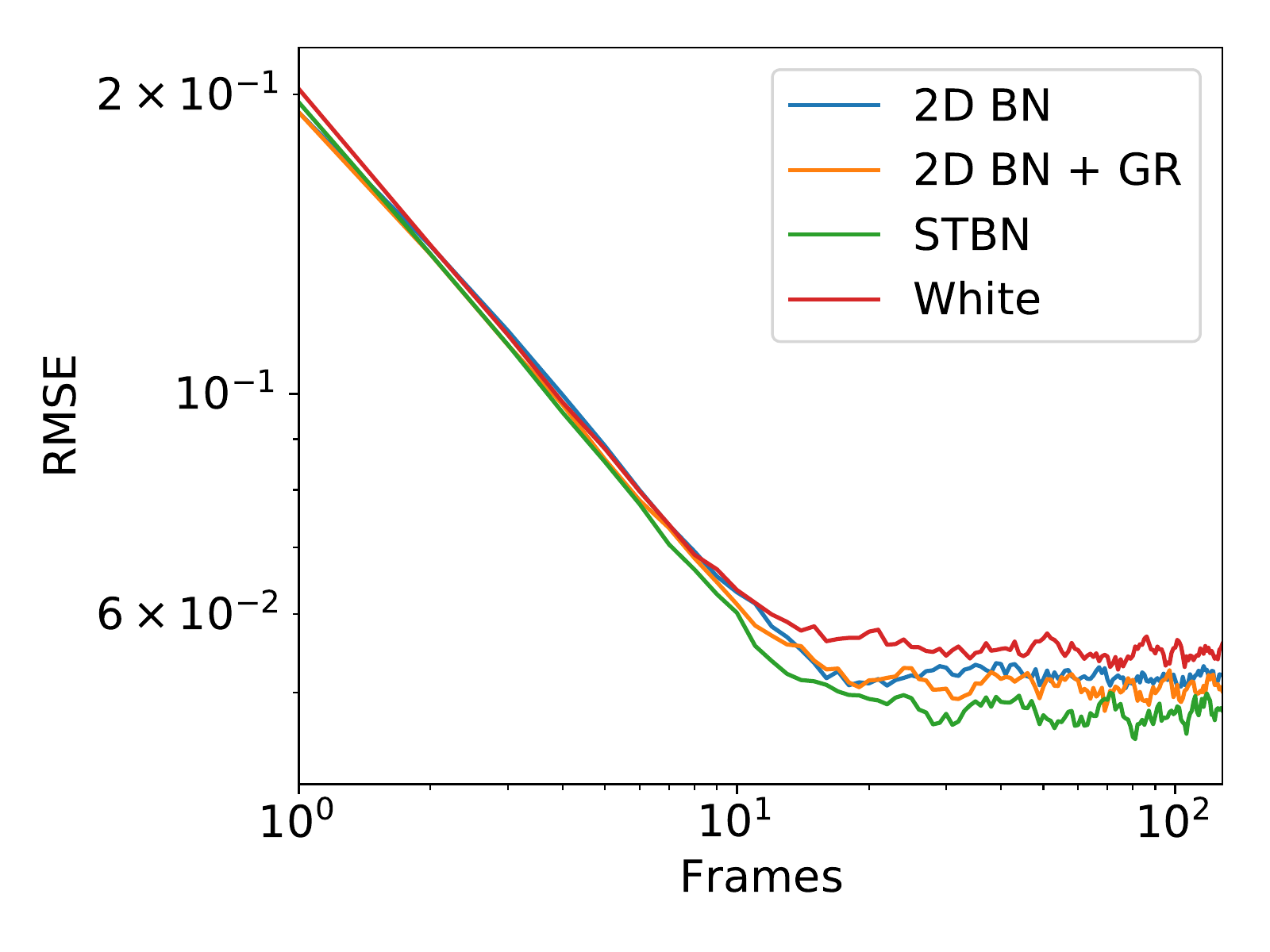} \\
		\end{tabular}
	}
	\caption{Convergence rate of ray traced AO. Under both Monte Carlo integration and TAA, STBN
	converges much better than the other types of noise.}
	\label{fig_ao_convergence}
\end{figure}

As shown in Fig.~\ref{fig_volume_renders}, when using white noise to sample the free-flight collision distance, we see undesired 
low frequency clusters in the image. This error pattern is greatly improved when using 2D blue noise masks,
and these low-frequency 
clusters are removed. However, over the course of several frames, the white noise exhibited by individual blue noise masks
over time results in poor convergence. Convergence over time is greatly improved by using the golden ratio LDS, but at a
compromise to the spatial blue noise frequency. Our spatiotemporal blue noise achieves competitive convergence to both
the golden ratio LDS and the LDS by Heitz and Belcour~\shortcite{Heitz2019}, as shown in Fig.~\ref{fig_volume_convergence}.
However, note that our results do not damage the spatial blue noise properties of these prior techniques.
Moreover, our method is more 
temporally stable with somewhat faster convergence under EMA, which can also be seen more clearly in the 
supplemental material.

In Fig.~\ref{fig_teaser}, we also report image 
error metrics, namely RMSE, SSIM, and \FLIP~\cite{Andersson2020}, against a
single scattering ground truth image. Although 2D blue noise already shows a favorable 
image error, our spatiotemporal blue noise further reduces this error, by $1.4\times$ for \FLIP and $1.8\times$ for 
SSIM.

\subsection{Ray Traced Ambient Occlusion}
\label{sec_applications_rtao}

Ray traced ambient occlusion requires 2D vectors for each AO sample, but our blue noise
masks only contain a scalar value per entry. As an imperfect solution to this, we use independent
spatiotemporal blue noise values for each $x$ and $y$ value of each sample taken per frame. For example, four
samples per pixel would require eight independent spatiotemporal blue noise sources.  Just as we
did for dithering in Section.~\ref{sec_applications_dithering}, we will use the R2 low discrepancy
sequence to generate a texture read offset for each independent value needed.

While there are more sophisticated AO algorithms, the one described here is meant for giving high-quality
results under low sample counts, such as 1--4 samples per pixel (SPP), or lower than 1~SPP if run at
less than full resolution.
The algorithm works by reading two independent spatiotemporal blue noise values per sample per pixel and
uses them to calculate a normal oriented hemispherical direction to shoot an AO ray in. The ray is limited to
a scene dependent maximum length value, and the AO shading amount is the percentage down the ray length that
a hit occured. When taking multiple samples, they are averaged together for a final AO value for
that pixel.

Rendered results are shown in Fig.~\ref{fig_ao_renders}, where it can be seen that
our spatiotemporal blue noise provides better image quality than
2D blue noise as in previous examples, but is also significantly better than the golden ratio.  The golden ratio fails
under this algorithm because for a specific pixel, the same location of the same blue noise texture is read
every frame, and the golden ratio multiplied by the frame number is added to both $x$ and $y$ to get the
sample.  This makes for a very poor two-dimensional low discrepancy sequence---using the same
irrational number for both axes. The convergence graphs shown in Fig.~\ref{fig_ao_convergence} convey
the same story.

\section{Conclusions and Future Work}
\label{sec_conclusions}

We have shown how simple modifications to the well known void and cluster algorithm can generate
spatiotemporal blue noise masks. We have shown how these blue noise masks can be useful for a variety
of low sample count rendering algorithms with the goal of getting desirable blue noise error patterns
while also converging faster than other methods which use blue noise masks.  Lastly, we showed how
these blue noise masks can be thresholded to take these properties into the realm of blue noise
sampling. 

The largest limitation is that these blue noise masks have scalar values per entry as opposed to
vector values.  Many rendering techniques require stochastic vector values to operate, and while
streams of scalar values show benefit in simpler algorithms like ambient occlusion or sampling
area lights, this can be more challenging for more complex algorithms such as path tracing.  The
problem can be alleviated by using our noise in techniques such as Heitz and Belcour~\shortcite{Heitz2019} 
(see Appendix A), but we
are also interested in seeing if our generalized blue noise masks could be extended to include vectors,
to get better results more directly, without the issues associated with approximately inverting
pixel rendering.  One way of doing this could be to apply the energy formulation we describe in this
paper to the technique of Georgiev and Fajardo~\shortcite{Georgiev2016} while extending that technique
to three dimensions.

Another limitation of these blue noise masks is that they are limited to blue noise.  While it is true
that 2D blue noise in screen space is a desirable property, it would be nice to be able to use other
sequences on the time axis without sacrificing quality for better convergence. Different
frequency characteristics may be desired for filtering reasons as well.

Importance sampling is a topic which is largely at odds with using specific sample patterns, but blue
noise itself happens to keep desirable properties better when put through warping functions as
pointed out by Pharr~\shortcite{Pharr2019}.  However, we would like to explore extending these blue
noise masks to allow axis groups to have non uniform distributions.  This would allow blue noise to
be generated in \emph{post-warp} space, meaning the blue noise would not be damaged in any way and
could have importance sampling baked into it. While some PDFs may be very specific---such as a HDRI
skybox image---other PDFs would get much more re-use, such as GGX for specular reflections.

Generated spatiotemporal blue noise masks and the code to generate them can be found at:

https://github.com/NVIDIAGameWorks/SpatiotemporalBlueNoiseSDK

A zip of the supplemental material can be found at:

https://drive.google.com/file/d/10FCzKR4qIKESz7AFh4-gKNlv1EjN5sDM

\bibliographystyle{ACM-Reference-Format}
\bibliography{references}

\clearpage
\appendix

\section{Frequency Analysis of Integration and Inversion}
\label{sec_freq_analysis}
\small
We develop a simple frequency analysis of integration with blue noise offsets, and explain the inversion method in Heitz and Belcour~\shortcite{Heitz2019}.
While our theory is closely inspired by previous work~\cite{Durand2011,Ramamoorthi2012,Pilleboue2015}, we do not believe this frequency analysis of the 
spatial distribution of error has appeared before, 
and it may provide valuable insight into previous algorithms.  However, it is not required for understanding our method in the main paper.

For simplicity, consider a 1D integral at a single pixel, 
\begin{equation}
	I_N = \int s(y) f(y)\, dy
	\ \ \ \ \ \ \ \ \ \ \ \ \
	s(y) = \frac{1}{N} \sum_{i=1}^{N} \delta(y-y_i),
\end{equation}
where $s(y)$ is the sampling pattern, $f(y)$ is the integrand we seek
to integrate by Monte Carlo, $I_N$ is the output image pixel radiance from using $N$ samples, and $y$ is the
variable we are integrating over.  Note that the sampling pattern $s$ is actually $N$
points $y_i$, which can be treated as delta functions.  
The integral can also be estimated in the Fourier domain~\cite{Durand2011}, 
\begin{equation}
	I_N = \left. \left(S(\omega) \otimes F(\omega)\right) \right|_{\omega = 0} =
	\int_{-\infty}^{\infty} S(\omega)F^{*}(\omega) \,d\omega,
	\label{eq:integration}
\end{equation}
where we use capital letters to denote the Fourier transforms, and $*$ denotes the complex conjugate (note that
$F^{*}(\omega) = F(-\omega)$).

We now consider perturbing the sampling pattern by a constant
$\gamma$.  That is, we replace all $y_i$ by $y_i + \gamma$.  In
practice, the values of $\gamma$ will differ at each pixel, in our
case with a blue noise-like pattern.  We now have, 
\begin{eqnarray}
	s(y;\gamma) =& \frac{1}{N}\sum_{i=1}^{N} \delta(y-y_i-\gamma) &= s(y-\gamma)\\
	S(y;\gamma) =& e^{-i \omega \gamma} S(y), &
\end{eqnarray}
where the last line follows simply from the addition theorem for
Fourier series.  Note
that the magnitude of the Fourier spectrum for the sampling pattern
remains the same, only its phase is shifted, with different phase
shifts at each pixel corresponding to each $\gamma$.
If we now plug the shifted sampling pattern for the integral we seek
into Equation~\ref{eq:integration}, we obtain, 
\begin{equation}
	I_N = \int S(\omega) F^{*}(\omega) e^{-i \omega \gamma}\, d\omega.
	\label{eq:shifted}
\end{equation}
Now, we can define $G(\omega) = S(\omega)F^{*}(\omega)$, where the
corresponding angular domain function is given by a convolution\footnote{
	Since $F^{*}(\omega) = F(-\omega)$ so that $G(\omega) = S(\omega) F(-\omega)$, this convolution has a 
	slightly non standard form with the plus sign instead of the conventional minus in the primal angular domain, $s(x)\otimes f(x) = \int s(x+y) f(y)\,dy$.
}
$g(x) =
f(x)\otimes s(x)$.  Note that Equation~\ref{eq:shifted} can be viewed as an inverse Fourier transform, 
evaluated at the value of $-\gamma$ (we omit normalizations),
\begin{eqnarray}
	g(x) = s \otimes f & = & \int \left(S(\omega) F^{*}(\omega) \right)
	e^{i \omega x} \,d\omega \\
	I_N & = & g(-\gamma).
\end{eqnarray}

\ignore{
There are a couple of useful sanity checks and intuitions related to
the above result.  If $\gamma = 0$, so we are considering the original
unshifted pattern, the above result reduces to
Equation~\ref{eq:integration}.  However, we can also consider this result
in the primal domain, 
\begin{equation}
	g(0) = \left. \left( f \otimes s \right)\right|_{0} = \int f(y)
	s(y)\,dy.
\end{equation}

Another noteworthy special case is what happens if the sampling
pattern is simply $s(y) = 1$ as in the ideal case.  In this case
$S(\omega)$ is nonzero and a delta function only for $\omega = 0$, and
we simply pull down $F(0)$, which is simply the desired integral $\int
f(y)\, dy$.  Note that in this case, the result is independent of
$\gamma$ as it should be if we have an ideal sampling function.  We
can also derive this result in the primal domain from the definition
of $g = f \otimes s$.  We can write more formally, $g(x) = \int
f(x+y)\, s(y)\,dy = \int f(x+y)\, dy = \int f(u)\, du$ which is 
the original integral if $s = 1$.  
}

Note that this analysis holds not just for the value $I_N$ but for the error as well, if we simply zero out the DC term in $S$
(the DC term corresponds to the ideal sampling pattern resulting in the true integral value).

\ignore{
Finally, note that in error analyses of sampling patterns, it is
common to subtract out the constant term, so that one can define
$\hat{s} = s - 1$.  Then $\hat{s}$ accounts for the error or noise,
corresponding to the deviation of the sampling pattern from the ideal
but unachievable unit sampling pattern.  In the frequency domain, it
corresponds to setting the $S(0)$ term to $0$ while keeping the other
frequencies, which contribute to the noise.  Now, if we are
considering blue noise, it is clear that the low-frequency components
of $F$ will not contribute to the error, since they will be nullified
by the blue noise floor.  This analysis is still in terms of the
integral rather than the spatial distribution.
}

Finally, we analyze the spatial distribution of the noise, which has not usually been considered in previous frequency analyses.
We follow the reasoning from Heitz and Belcour~\shortcite{Heitz2019}, which assumes the function $f$ (and hence $g$) is locally constant in a patch. 
From the above results, we can then define, 
\begin{equation}
	I_N = g(-\gamma) \ \ \ \ \ \Rightarrow \ \ \ \  I_N(x) = g(-\gamma(x)).
	\label{eq:bgamma}
\end{equation}
\begin{figure}
	\centering
	\setlength{\fboxsep}{0pt}%
	\setlength{\fboxrule}{0.0mm}%
	\setlength{\tabcolsep}{1.5pt}%
	\renewcommand{\arraystretch}{0.95}
	\newcommand{\imgW}{0.8\columnwidth}
	\newcommand{\imgWSm}{0.15\columnwidth}
	{\scriptsize 
		\begin{overpic}[width=\imgW]{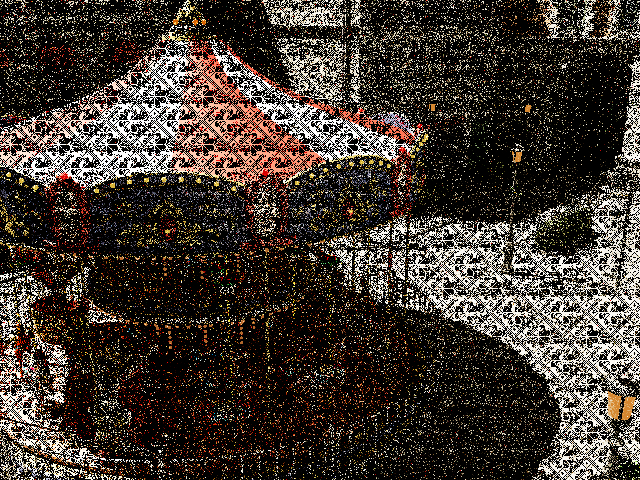}
			\put(3,3){\includegraphics[width=\imgWSm]{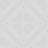}}
		\end{overpic}
	}
	\caption{One SPP path tracing, rearranging the seeds using the Heitz/Belcour technique to make the
	render noise match the stylized texture shown in the lower left.}
	\label{fig_hb_renders}
\end{figure}

\begin{figure*}
	\centering
	\begin{subfigure}[b]{.24\textwidth}
		\includegraphics[width=\textwidth]{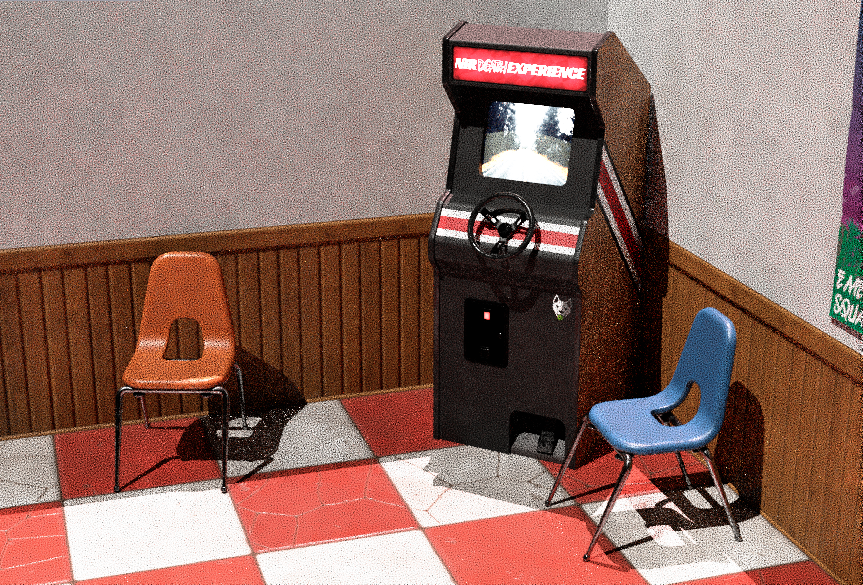}
		\caption{Heitz and Belcour using STBN}
	\end{subfigure}
	\begin{subfigure}[b]{.24\textwidth}
		\includegraphics[width=\textwidth]{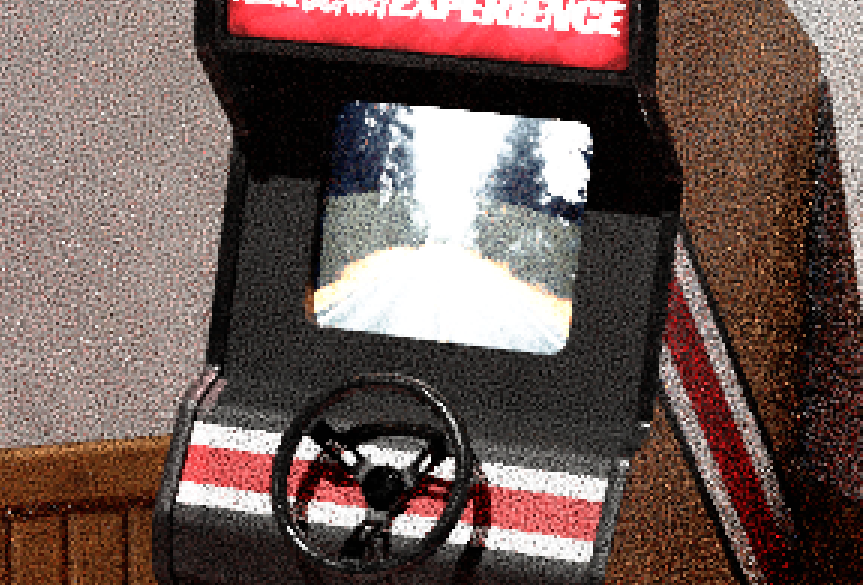}
		\caption{2D BN}
	\end{subfigure}
	\begin{subfigure}[b]{.24\textwidth}
		\includegraphics[width=\textwidth]{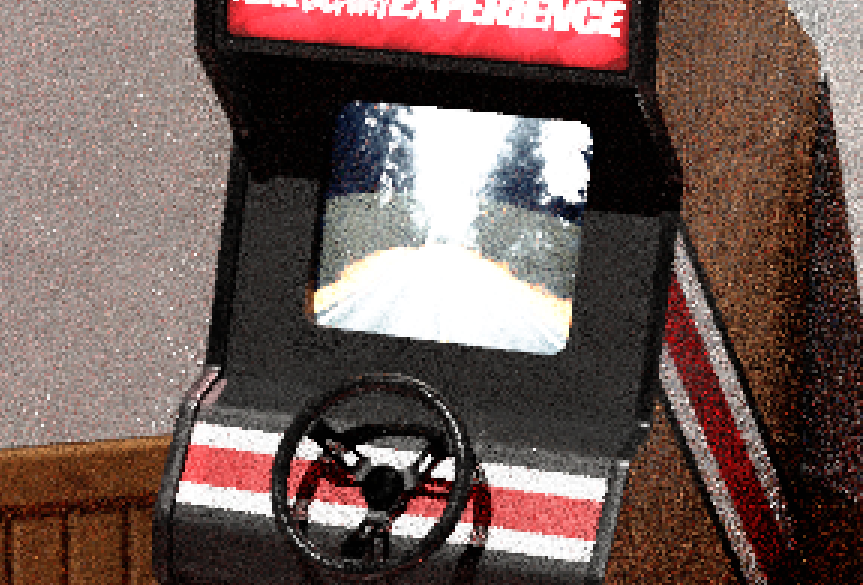}
		\caption{STBN (Ours)}
	\end{subfigure}
	\begin{subfigure}[b]{.24\textwidth}
		\includegraphics[width=\textwidth]{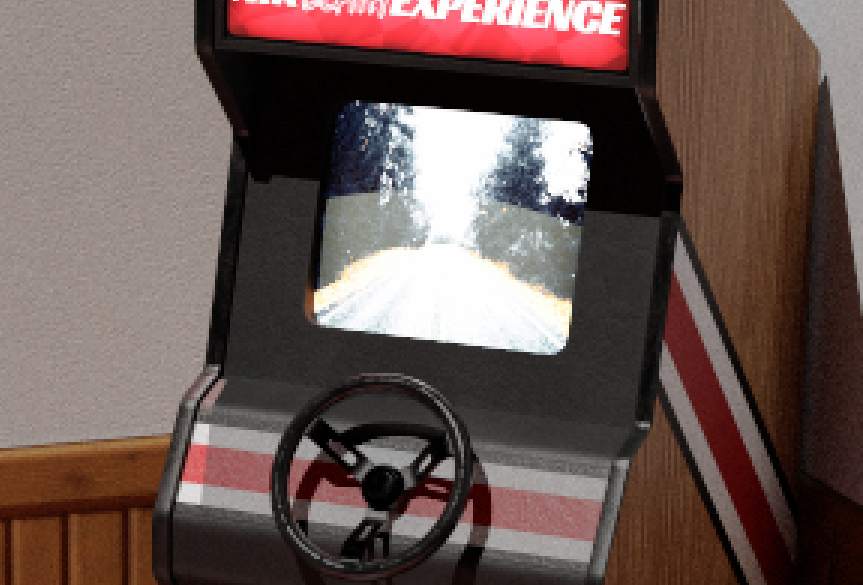}
		\caption{Ground Truth}
	\end{subfigure}
	\caption{A comparison of blue noise versus spatiotemporal blue noise (center) applied to the 
	seed rearranging technique by Heitz/Belcour, with a ground truth image on the right. 
	Both noise signals were filtered using an exponential moving average and $\alpha=0.1$. 
	This example illustrates that our spatiotemporal blue noise can be used in combination 
	with state of the art techniques such as the Heitz/Belcour algorithm. } 
	\label{fig_hb_stbn}
\end{figure*}

In Heitz and Belcour's work~\shortcite{Heitz2019} and in this paper, we want the error in $I_N(x)$ to be distributed as a blue noise pattern.  The simplest approach~\cite{Georgiev2016}
is to choose the offsets/seeds $\gamma(x)$ in a blue-noise pattern.  However, the pattern in $I_N$ is not necessarily blue-noise since it is transformed
by the function $g(-\gamma(x))$.  Instead, if we want the distribution $I_N(x)$ (or equivalently the error in the distribution) to achieve some desired pattern $I_N(x) \sim \alpha(x)$, then 
we seek that $g(-\gamma(x)) = \alpha(x)$ where
$\alpha(x)$ is the desired blue noise pattern,  
\begin{equation}
	\gamma(x) = g^{-1}(\alpha(x)),
	\label{eq:gammag}
\end{equation}
where we have ignored the negative sign, and we need to explicitly invert the function $g$.  However, this seems
intractable, since we do not know $g$ nor its inverse (or even the 
integral of $f$); 
indeed that's what we are trying to find by Monte Carlo integration.  
Instead, Heitz and Belcour~\shortcite{Heitz2019} perform this inversion in an ordinal numerical fashion within a local patch.
They sort the seeds $\gamma(x)$ and the function
values $g(\gamma(x))$ from just applying naive blue-noise patterns
in the first step.  Then, instead of doing exact inversion and
interpolation, one can just map the function ordinally in terms of the
sorted order.  So now, in the next frame, if you have a seed
$\alpha(x)$ at pixel $x$, we get the corresponding ordinal value of
$\alpha$ and see which $g(\gamma)$ gave that corresponding ordinal
value, effectively inverting $\gamma(x) = g^{-1}(\alpha(x))$.


Note that the inversion method is orthogonal to the specific desired pattern $\alpha(x)$ (or for that matter the initial pattern $\gamma(x)$).
Indeed, we have successfully used the method even to produce stylized error patterns as can be seen
in Fig.~\ref{fig_hb_renders}.

We have integrated our spatiotemporal blue noise into Heitz and Belcour's method~\shortcite{Heitz2019},
using both seed sorting and retargeting and show improved convergence performance, as shown in Fig.~\ref{fig_hb_stbn}. 


\section{Higher Dimensional Blue Noise Masks}
\label{higher_dimensional_masks}
\small

\ignore{
Spatiotemporal blue noise masks are valuable in any animated situation where 2D blue noise textures
are currently used, as they are a solution to the problem of animating blue noise masks.  Where other methods
either damage the blue noise spatially, or are white noise over time, spatiotemporal blue noise has
both a high quality blue noise spectrum over space, while also having a blue noise spectrum over time.

While there are many algorithms that fall into this usage case of only needing a single scalar per
pixel per frame, there are many algorithms that need more than that. At the extreme end is general
path tracing which can need a large number of random values per sample, and may take several samples
per frame.  Even before reaching path tracing however, techniques such as ray traced area lights or
raytraced Ambient Occlusion already would like at minimum two values per pixel.

To aid this,}
Our generalized void and cluster algorithm runs in $D$ dimensions to generate a mask $M$, where $D$ is the number of parameters needed to
index into the mask to get a scalar value.  These parameters are the axes of the blue noise mask.
The $D$ dimensions are broken up into one or more sets $G$, where each set of $G$ contains one or
more dimensions. A specific set $g$ of $G$ with a membership count of $d$ implies that all $d$ dimensional
projections of the $D$ dimensional blue noise mask should be $d$ dimensional blue noise, when only the
axes within that group vary, and all other axes are held constant. We will also define a group of
axes $h$ as being all axes which are not in $g$.

Once the dimensions are grouped, each group $g$ is naturally mapped to an energy function $E_g$ by
only using the dimensions present in the group (denoted as $\mathbf p_g$ and $\mathbf q_g$) within
the usual Gaussian energy function, so long as the axes from the corresponding group $h$ are equal
between the two pixels. The energy function between two pixels is the sum of all $E_g$ functions
between those pixels, and the energy field $F$ is the sum of energy at each pixel, from every other pixel.
This is summarized as
\begin{equation}
\label{eqn_generalized_energy}
	\begin{array}{l}
	E_g(\mathbf p,\mathbf q) =
	\begin{cases}
	\exp{\left(-\frac{\Vert \mathbf p_g-\mathbf q_g \Vert ^2}{2\sigma_g^2}\right)}, & \text{if\ } p_h = q_h \\ 
	0, &\text{otherwise.} \\
	\end{cases}
	\vspace*{1mm}
	\\
	E(\mathbf p,\mathbf q) = \sum_{\mathbf g \in G} E_g(\mathbf p,\mathbf q), \vspace*{1mm}\\
	F(\mathbf p) = \sum_{\mathbf q \in M} E(\mathbf p,\mathbf q).
	\end{array}
\end{equation}
Each dimension can be of different size, can use a different $\sigma$ values to control
the frequency content of the result, and can also choose to compute distances toroidally or not.
The original void and cluster algorithm can be seen as a special case such that $D$ is any arbitrary value, and that there
is only a single group in $G$ which contains all axes.  Thus, the void and cluster algorithm
makes $D$ dimensional blue noise masks.
When considering spatiotemporal blue noise, $D$ is 3, and $G$ has two groups in it: $g_{xy}$ and $g_z$.
In that sense, spatiotemporal blue noise can also be seen as a 2D$\times$1D blue noise mask.

\begin{figure}[tb]
	\centering
	\setlength{\fboxsep}{0pt}%
	\setlength{\fboxrule}{0.0mm}%
	\setlength{\tabcolsep}{1.5pt}%
	\renewcommand{\arraystretch}{0.95}
	\newcommand{\imgW}{0.28\columnwidth}
	\newcommand{\imgWSmall}{0.07\columnwidth}
	{\scriptsize 
		\begin{tabular}{l|cccccc}
			& XY & ZX & WX & ZY & WY & ZW \\
			\hline
			\rotatebox{90}{2D$\times$2D} &
			\includegraphics[width=\imgW]{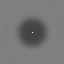} &
			\rotatebox{90}{\includegraphics[width=\imgW]{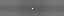}} &
			\rotatebox{90}{\includegraphics[width=\imgW]{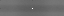}} &
			\rotatebox{90}{\includegraphics[width=\imgW]{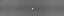}} &
			\rotatebox{90}{\includegraphics[width=\imgW]{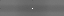}} &
			\includegraphics[width=\imgW]{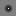} \\
			\hline
			\rotatebox{90}{2D$\times$1D$\times$1D} &
			\includegraphics[width=\imgW]{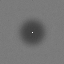} &
			\rotatebox{90}{\includegraphics[width=\imgW]{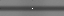}} &
			\rotatebox{90}{\includegraphics[width=\imgW]{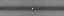}} &
			\rotatebox{90}{\includegraphics[width=\imgW]{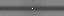}} &
			\rotatebox{90}{\includegraphics[width=\imgW]{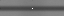}} &
			\includegraphics[width=\imgW]{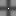} \\
			\hline
		\end{tabular}
	}
	\caption{DFTs of the 2D projections of 4D blue noise masks that are $64\times 64\times 16\times 16$. All
	projections averaged to show expected frequency spectra.}
	\label{fig_2dx1dx1d_dft}
\end{figure}

Frequency analysis of a 2D$\times$1D$\times$1D and 2D$\times$2D mask can be seen in Figure~\ref{fig_2dx1dx1d_dft}, which
shows the desired frequency behaviors for axis pairs. The 2D$\times$2D blue noise shows 2D blue noise on the XY plane
and the ZW plane, but white noise everywhere else.  The 2D$\times$1D$\times$1D blue noise shows 2D blue noise on
the XY plane and shows 1D blue noise on both the Z and W axes.

\begin{table}
	\centering
	\caption{The storage size of masks, and generation time.}
	\begin{tabular}{|l|c|c|}
		\hline
		Dimensions & Size & Generation Time \\
		\hline
		$64\times 64$ & 4~kB & < 1 second \\
		$32\times 32\times 16$ & 16~kB & < 1 second \\
		$32\times 32\times 32$ & 32~kB & 2 seconds \\
		$64\times 64\times 16$ & 64~kB & 10 seconds\\
		$256\times 256$ & 64~kB & 10 seconds \\
		$128\times 128\times 8$ & 128~kB & 44 seconds \\
		$64\times 64\times 64$ & 256~kB & 3 minutes\\
		$128\times 128\times 32$ & 512~kB & 12 minutes\\
		$64\times 64\times 16\times 16$ & 1~MB & 48 minutes\\
		$64\times 64\times 64\times 64$ & 16~MB & 206 hours (Estimated)\\
		\hline
	\end{tabular}
	\label{mask_size_gentime}
\end{table}

Generation time of blue noise masks is a function of the total pixel count $n$,
and is nearly $O(n^2)$ so that doubling the pixel count roughly quadruples the generation time. 
Table~\ref{mask_size_gentime} shows some reasonable mask sizes, their size in bytes assuming a
single channel 8-bit texture, and the time taken to generate them. We have found that smaller textures such
as $64\times 64\times 16$ (64~kB) for spatiotemporal blue noise, and $64\times 64\times 16\times 16$ (1~MB)
have been sufficient in our rendering tests.

Note that this generation time is a preprocess, and is simply used to generate a blue noise texture.  At 
run-time only a simple texture-read needs to be performed, so our algorithm (spatiotemporal blue noise 
masks in the main paper or higher-dimensional masks discussed here) has essentially no overhead, and can be 
included in any real-time image synthesis method.

\clearpage

\end{document}